\documentclass{cernyrep}

\usepackage{mathptmx}       
\usepackage{helvet}         
\usepackage{courier}        
\usepackage{type1cm}        
\usepackage{makeidx}         
\usepackage{graphicx}        
\usepackage{multicol}        
\usepackage[bottom]{footmisc}
\usepackage{color}
\usepackage{url}
\usepackage{varwidth}
\usepackage{xcolor}

\makeindex             

\usepackage{amsbsy}

\begin{document}

\title{Ion Acceleration---Target Normal Sheath Acceleration\thanks{Re-published, with permission, from P. McKenna et al (eds.), "Laser-Plasma Interactions and Applications", Scottish Graduate Series (Springer International Publishing, 2013).}}

\author{M. Roth and M. Schollmeier}

\institute{Institute for Nuclear Physics, Technische Universit\"{a}t Darmstadt, Darmstadt, Germany}

\maketitle

\begin{abstract}
Energetic ions have been observed since the very first laser-plasma ex\-peri\-ments. Their origin was found to be the charge separation of electrons heated by the laser, which transfers energy to the ions accelerated in the field. The advent of ultra-intense lasers with pulse lengths in the femtosecond regime re\-sulted in the discovery of very energetic ions with characteristics quite different from those driven by long-pulse lasers. Discovered in the late 1990s, these ion beams have become the focus of intense research worldwide, because of their unique properties and high particle numbers. Based on their non-isotropic, beam-like behaviour, which is always perpendicular to the emitting surface, the acceleration mechanism is called tar\-get normal sheath acceleration (TNSA). We address the physics of the mechanism and its dependence on laser and target par\-ameters. Techniques to explore and diagnose the beams, to make them useful for applications, are also addressed.
\end{abstract}

\section{Introduction}
\label{sec:intro}
Since the first irradiation of a target by a laser, the generation of energetic ions has been well known. The origin of those ions is the electric field generated by the charge separation as a result of the en\-ergy transferred from a long-pulse laser to the electrons, and their respective temperature \cite{gitomer}. The ions are then accelerated in the double-layer potential and can reach significant particle energies, expanding isotropically in all directions from the target front surface.
Since the advent of ultra-short-pulse lasers with pulse lengths of less than picoseconds, one of the most exciting results obtained in experiments using solid targets is the discovery of very energetic, very intense bursts of ions coming off the rear, non-irradiated surface in a very high quality, beam-like fashion. At the turn of the century, a number of experiments have resulted in proton beams with energies of up to several tens of megaelectronvolts generated behind thin foils irradiated by lasers exceeding hundreds of terawatts \cite{clark2000, snavely2000, maksimchuk2000}. Since the first observations, an extraordinary amount of experimental and theoretical work has been devoted to the study of these beams' characteristics and production mechanisms. Particular attention has been devoted to the exceptional accelerator-like spatial quality of the beams, and current research focuses on their optimization for use in a number of groundbreaking applications, addressed in Section \ref{sec:Applications}. But first we will focus on the best understood of all the possible acceleration mechanisms, so-called target normal sheath acceleration (TNSA).

The greater part of this chapter is drawn from Ref.~\cite{Schollmeierdiss}. Review articles about TNSA, the diag\-nostics of short-pulse laser plasmas, and applications in fast ignition can also be found in Refs. \cite{borghesi2006, roth2011, roth2002, logan2006, roth2009}.

\section{Target normal sheath acceleration: the mechanism}
\label{sec:TNSA - The mechanism}

\subsection{Initial conditions}
\label{subsec:Initial Conditions}

The primary interaction of a high-intensity short laser pulse with a solid target strongly depends on the contrast of the laser pulse, that is, the ratio of unwanted, preceding laser light to the main pulse. At peak intensities exceeding \(10^{20}\UW/\UcmZ^2\) even a contrast of \(10^6 \) is sufficient to excite a plasma that is expanding towards the incoming main pulse. As a common source of this unwanted laser light, amplified spontaneous emission or prepulses, caused by a limited polarization separation in regenerative amplifiers, have been identified. This ablative plasma sets the stage for a wealth of uncontrolled phenomena at the interaction of the main pulse with the target. The laser beam can undergo self-focusing due to ponderomotive force or relativistic effects, thereby strongly increasing the resulting intensity, or it can break up into multiple filaments, or, finally, it can excite instabilities that ultimately lead to the production of energetic electrons.
Moreover, the ablative pressure of blow-off plasma caused by the incident laser energy prior to the main pulse launches a shockwave into the target, which can ultimately destroy the target before the arrival of the main pulse. We address this issue, even though it is not directly related to the TNSA mechanism, because of its influence on the electron spectrum and the thickness of targets that can be used in practice.

\subsection{General description}
\label{subsec:General Description}

\begin{figure}
        \centering
        \includegraphics[width=0.9\textwidth]{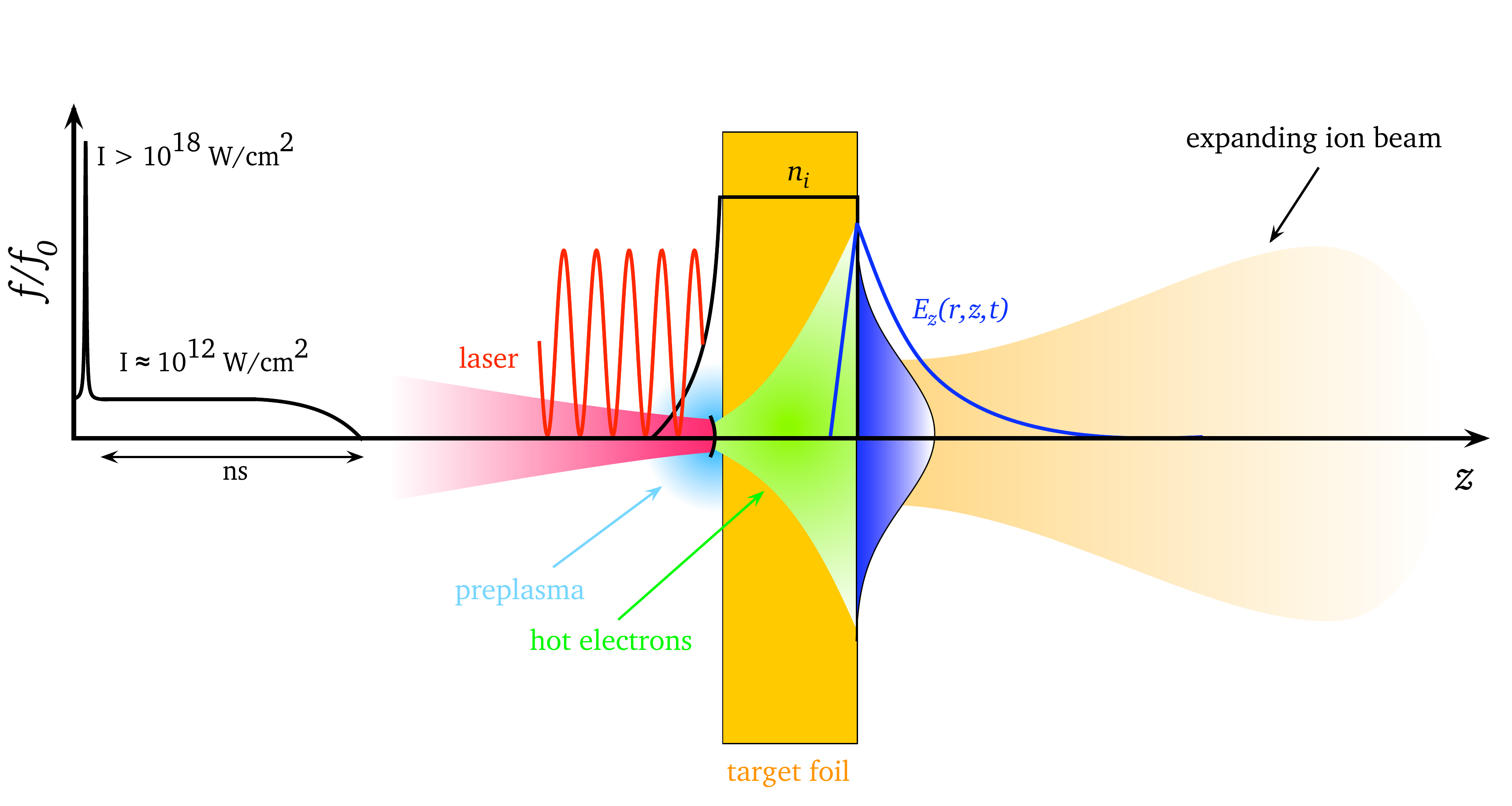}
                \caption{Target normal sheath acceleration. A thin target foil with thickness \(d = 5\)--\(50\Uum\) is irradiated by an intense laser pulse. The laser prepulse creates a preplasma on the target's front side. The main pulse interacts with the plasma and accelerates megaelectronvolt electrons, mainly in the forward direction. The electrons propagate through the target, where collisions with the background material can increase the divergence of the electron current. The electrons leave the rear side, resulting in a dense sheath. An electric field due to charge separation is created. The field is of the order of the laser electric field (\(\sim\UVZ[T]/\UmZ\)), and ionizes atoms at the surface. The ions are then accelerated in this sheath field, pointing in the target normal direction.}
                \label{tnsa-scheme}
\end{figure}

Before going into detail, it is worth taking a step back and qualitatively looking at the general picture of the TNSA ion acceleration mechanism. Let us interpret the process of generating a proton beam by TNSA as a new variation on a familiar theme---acceleration by a sheath electrostatic field generated by the hot-electron component. We assume the interaction of an intense laser pulse well exceeding \(10^{18}\UW/\UcmZ^2 \) with a solid thin foil target as the standard case for TNSA. The interaction of the intense laser pulse with the preformed plasma and the underlying solid target constitutes a source of hot elec\-trons with an energy spectrum related to the laser intensity. This cloud of hot electrons penetrates the foil at, as we shall see, an opening angle of about \(30^{\circ}\) and escapes into the vacuum behind the target. The target's capacitance, however, allows only a small fraction of the electrons to escape before the target is sufficiently charged that escape is impossible for even megaelectronvolt electrons. Those electrons are then electrostatically confined to the target and circulate back and forth through the target, laterally expanding and forming a charge-separation field on both sides over a Debye length. At the rear surface there is no screening plasma, owing to the short time-scales involved, and the induced electric fields are of the order of several teravolts per metre. Such fields can ionize atoms and rapidly accelerate ions normal to the initially unperturbed surface. The resulting ion trajectories thus depend on the local orientation of the rear surface and the electric field lines driven by the time-dependent electron density distribution. As the ions start from a cold solid surface just driven by quasi-static electric fields, the resulting beam quality is extremely high, as we shall see. This process is illustrated in \Fref{tnsa-scheme}.

\subsection{Electron driver}
\label{subsec:Electron driver}

Current laser systems are not yet capable of accelerating ions directly. Therefore, all existing laser ion acceleration mechanisms rely on the driving electron component and the resulting strong electric field caused by charge separation. Thus the electron driver is extremely important, and will be discussed here in detail.
As a rule of thumb, particle-in-cell calculations \cite{wilks1999, wilks1993, lasinski1999} have indicated that the so-called hot-electron component has a logarithmic-slope temperature that is roughly equal to the ponderomotive potential of the laser beam. This is represented by the cycle-averaged kinetic energy of an electron oscillating in the laser electromagnetic field, \(T_{\textrm{hot}} \approx U_{\textrm{pond} } \approx 1\UMeV \times (I \lambda^2/10^{19}\UW\cdot\Uum^2/\UcmZ^2)^{1/2} \) in the relativistic regime \cite{wilks2000}. The relativistic electrons are directed mainly in the forward direction \cite{Norreys1999}; hence, the particle distribution function can be simplified by a one-dimensional Maxwell--J\"uttner distribution, which is close to an ordinary Boltzmann distribution.

The conversion efficiency from laser energy to hot electrons is not perfect, and only a fraction \(\eta\) is converted. The total number of electrons is
\begin{equation}
n_0=\frac{\eta E_\textrm{L}}{c \tau_\textrm{L} \pi r_0^2 k_\textrm{B}T_{\textrm{hot}}}~,
\label{eq:total_el}
\end{equation}
following  a scaling with intensity as
\begin{equation}
\eta=1.2 \times 10^{-15} I^{0.74}~,
\label{eq:frac}
\end{equation}
with the intensity in  watts per square centimetre reaching up to 50\% \cite{fuchs2006}.
For ultra-high intensities, \(\eta\) can reach up to 60\% for near-normal incidence and up to 90\% for irradiation under \(45^\circ\) \cite{ping2008}. A discussion on which distribution function best fits the experimental data is given in Ref. \cite{key1998} and, in more detail, in Ref. \cite{Davies2002}. However, neither theoretical nor experimental data give a clear answer to the question about the shape of the distribution function.

Given the intensities of modern short-pulse lasers, therefore, copious amounts of energetic elec\-trons are generated and, in contrast with thermal electrons in long-pulse laser plasmas, are pushed into the target. It is fair to estimate that a fraction of \(N=\eta E_\textrm{L}/k_\textrm{B}T_{\textrm{hot}}\) electrons in the megaelectronvolt range are created, where \(E_\textrm{L}\) is the laser energy.   These electrons have typical energies such that their mean free path is much longer than the thickness of the targets typically used in experiments. While the electrons propagate through the target, they constitute a current that exceeds the Alfv\'{e}n limit by several orders of magnitude. Alfv\'{e}n found that the main limiting factor on the propagation of an electron beam in a conductor is the self-generated magnetic field, which bends the electrons back towards the source \cite{alfven}. For parameters of relevance for inertial confinement fusion, a good review is given in Ref. \cite{davis2004}. So as not to exceed the limit of \( j_\textrm{A} = m_\textrm{e}c^3 \beta \gamma/e = 17 \beta \gamma~(\UkAZ)\), the net current must be largely compensated for by return currents, to minimize the resulting magnetic field. The return currents will be driven by the charge separation in the laser--plasma interaction region and strongly depends on the electrical conductivity of the target, as those currents are lower in energy and thereby affected by the material properties. The large counter-streaming currents also give rise to instabilities, which affect the forward motion of the electrons. The influence of limited electrical conductivity on the inhibition of fast-electron propagation has been addressed in Ref.\cite{batani2002},  also with respect to space charge separation. Without the return currents, the electric field would stop the electrons in a distance of less than 1\Unm \cite{bell2006}. The electric field driving the return current, in turn, can be strong enough to stop the fast electrons. This effect, known as transport inhibition, is prominent in insulators, but almost negligible in conductors \cite{pisani2000}.

\begin{figure}
        \centering
        \includegraphics[width=0.8\textwidth]{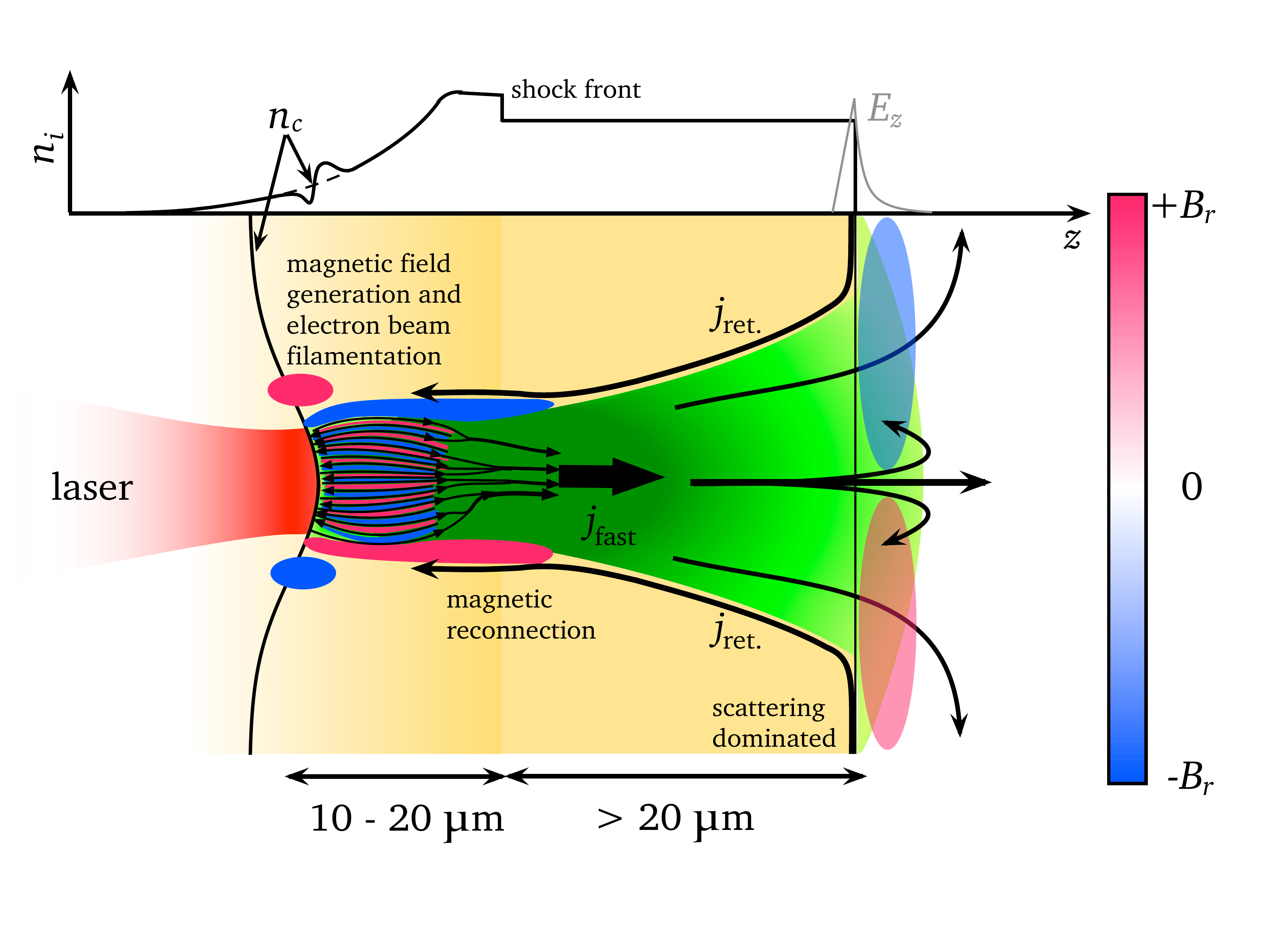}
                \caption{Schematic of laser-generated fast-electron transport. The laser (shown in red) impinges on a preplasma with exponential density profile from the left side. The light pressure leads to profile steepening, depicted in the graph at the top of the figure. An ablation plasma creates an inward-travelling shockwave that heats, ionizes, and compresses the target. Fast electrons are created by the laser, propagating into the dense plasma towards the target's rear side. The high electron current \(j_{\textrm{fast}}\) can lead to filamentation and magnetic field generation (shown by the light red- and blue-coloured areas), as well as driving a return current \(j_{\textrm{ret}}\). The global magnetic field tends to pinch the fast-electron current. Electrons propagating in the dense solid matter interact with the background material by binary collisions. This leads to a spatial broadening of the electron distribution, which becomes the major effect for longer distances. At the rear side, the electrons form a sheath and build up an electrostatic field \(E_z\) (grey line in graph). This can lead to refluxing (recirculation) of the electrons, heating the target even further.}
                \label{etransport}
\end{figure}

The propagation of electrons through the target is still an active field of research. As depicted in \Fref{etransport}, the laser pushes the critical surface \(n_\textrm{c}\), leading to a steepening of the electron density profile. The motion of the ablated plasma causes a shockwave to be launched into the target, leading to ionization and therefore a modification of the initial electrical conductivity. As soon as the electrons penetrate the cold solid region, binary collisions (multiple small-angle scattering) with the background material are no longer negligible. These tend to broaden the electron distribution, counteracting the magnetic field effect \cite{honrubia2006}. For long propagation distances (\(z \geq 15\Uum)\), the current density is low enough that broadening due to small-angle scattering becomes the dominating mechanism \cite{santos2007}.

The majority of data show a divergent electron transport. The transport full-cone angle of the electron distribution was determined to be dependent on laser energy and intensity, as well as target thickness. For rather thick targets \((d > 40\Uum) \) this value is around \(30^\circ \) for full width at half maximum (FWHM), whereas for thin targets \( (d \leq 10\Uum) \) published values are of the order of \(16^\circ \) (this figure was indirectly obtained by a fit to proton energy measurements) and are \( {\approx}150^\circ\) at most \cite{stephens2004}. It has been shown that different diagnostics lead to different electron transport cone angles \cite{lancaster2007}, so the nature of the `true' cone angle dependence with laser and target parameters remains unclear.

When the electrons reach the rear side of the target, they form a dense charge-separation sheath. The outflowing electrons lead to a toroidal magnetic field \(B_\theta\), which can spread the electrons over large transverse distances by a purely kinematic \(E \times B_\theta\) force \cite{sentoku2007}, sometimes called the fountain effect \cite{Pukhov2001}. The electric field created by the electron sheath is sufficiently strong to deflect   electrons back into the target, which then recirculate. Experimental evidence for recirculating electrons is  presented in Refs. \cite{ping2008, santos2002, nishimura2005}. The relevance to proton acceleration was first demonstrated by MacKinnon \textit{et al.} \cite{MacKinnon2002}, who measured a strong enhancement of the maximum proton energy for foils thinner than \(10\Uum\), compared with thicker foils. With the help of computer simulations, this energy enhancement was attributed to an enhanced sheath density caused by refluxing electrons. Further evidence of refluxing electrons was also found in an experiment discussed in Ref. \cite{Schollmeier2008}.

\begin{figure}
        \centering
        \includegraphics[width=0.8\textwidth]{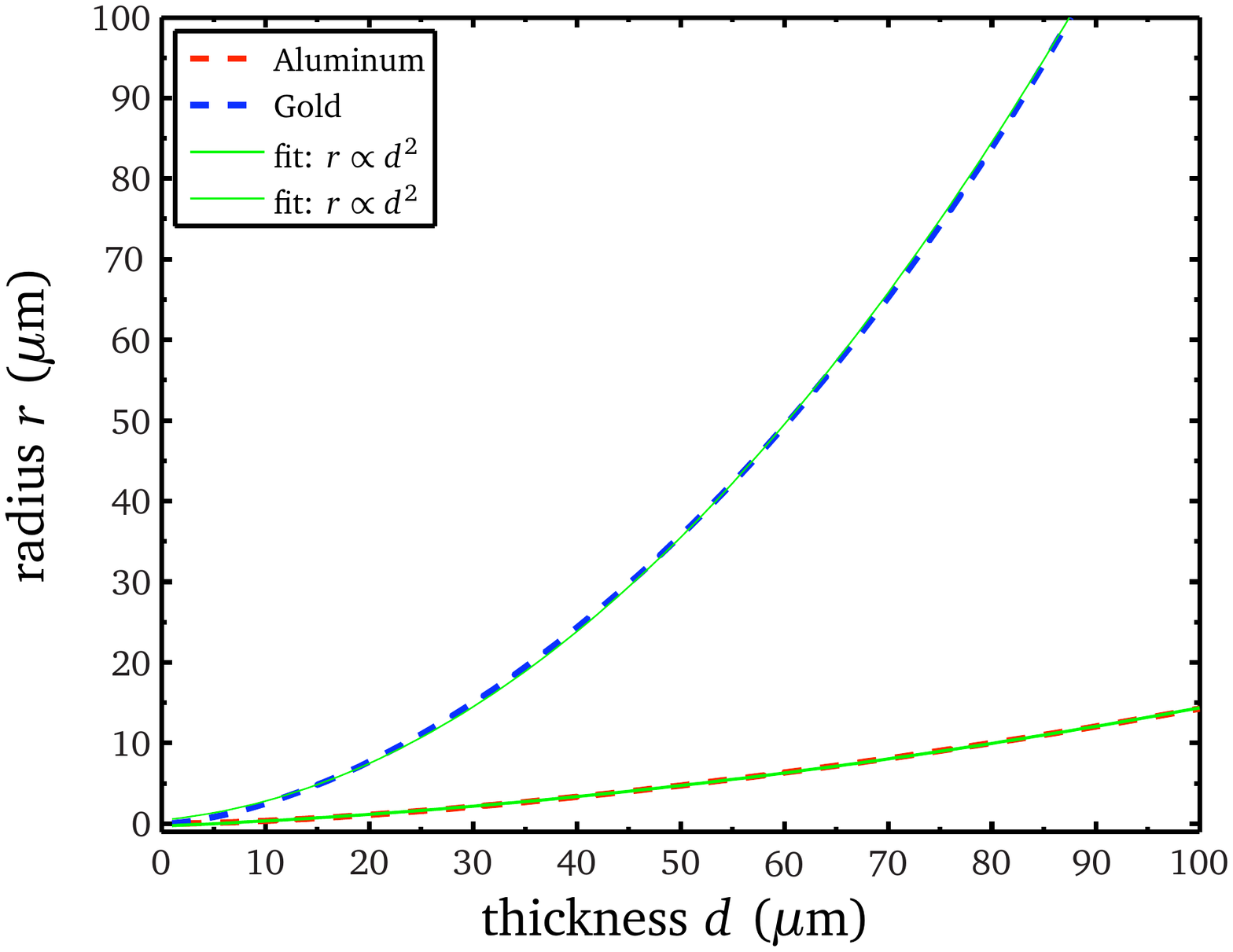}
                \caption{Increase in radius \(r\) of an electron distribution with target thickness \(d\). The calculation utilized \Eref{eq:broaden}, taking an energy of \(k_\textrm{B} T_{\textrm{hot}} \approx 1\UMeV\), corresponding to a laser intensity \(I = 10^{19}\UW/\UcmZ^2\).}
                \label{broaden}
\end{figure}

Neglecting the complicated interaction for thicknesses below \(d \approx 15\Uum\), a reasonable estimate for the electron beam divergence is the assumption that the electrons are generated in a region of the size of the laser focus and are purely collisionally transported to the rear side of the target. This is in agreement with most published data. The broadening of the distribution is then due to multiple Coulomb small-angle scattering, given analytically, \eg by Moli\`{e}re's theory in Bethe's description \cite{Bethe1953}.
To lowest order, the angular broadening, \(f (\Theta )\), follows a Gaussian (see Ref. \cite{Bethe1953}, Eq.  (27)),
\begin{equation}
f(\theta)=\frac{2\textrm{e}^{-\vartheta^2}}{\chi_\textrm{c}^2B} \sqrt{\theta/\sin\theta},
\label{eq:broaden}
\end{equation}
where the second term on the right-hand side is a correction for larger angles (from Ref. \cite{Bethe1953}, Eq. (58)). The angle \(\vartheta\) can be related to \(\theta\) by \(\vartheta=\theta/\chi_\textrm{c}B^{1/2}\). The transcendental equation, \(B - \textrm{ln}B = \textrm{ln} \left(\chi_\textrm{c}^2/\chi_{a'}^2\right)\), determines \(B\). The screening angle \(\chi_{a'}^2\) is given by \(\chi_{a'}^2 = 1.167 (1.13 + 3.76\alpha^2) \lambda^2/a^2\), where \(\lambda = \hbar/p\) is the de Broglie wavelength of the electron and \(a=0.885 a_\textrm{B} Z^{-1/3}\), with the Bohr radius \(a_\textrm{B}\). \(\alpha\) is determined by \(\alpha = Ze^2/(4\pi\epsilon_0\hbar\beta c)\) with the nuclear charge \(Z\), electron charge \(e\), and \(\beta = v/c\), where \(\epsilon_0\), \(\hbar\), and \(c\) denote the usual constants. The variable \(\chi_\textrm{c}\) is given by
\begin{equation}
\chi_\textrm{c}^2 = \frac{e^4}{4 \pi \epsilon_0^2c^2} \frac{ Z(Z+1)Nd}{\beta^2 p^2}~,
\end{equation}
with the electron momentum \(p\) and \(N = N_\textrm{A} \rho/A\) being the number of scattering atoms, determined by Avogadro's number \(N_\textrm{A}\), material density \(\rho\), and mass number \(A\). \(\chi_\textrm{c}\) is proportional to the material thickness \(d\) and density \(\rho\) as \(\chi_\textrm{c} \propto (\rho d)^{1/2}\). Since \(\chi_\textrm{c}\) determines the width of \( f (\theta)\), the angular spread of the electron distribution propagating through matter is proportional to its thickness as well as its density.
The analytical formula allows us to estimate the broadening of the laser-accelerated electron distribution during the transport through the cold solid target. For a laser intensity \(I_\textrm{L} = 10^{19}\UW/\UcmZ^2\), the mean energy (temperature) is \(k_\textrm{B} T_{\textrm{hot}} \approx 1\UMeV\). The increase in distribution radius \(r\) with target thickness \(d\) is shown in \Fref{broaden}. The electrons were propagated in aluminium (red dashed lines) and gold (blue dashed lines). Aluminium does not lead to a strong broadening, owing to its low density and \(Z\); compare this with the broadening observed in gold. The graph shows that in each case the radius at the rear side scales as \(r \propto d^2\) (green lines).

The estimate based on an electron distribution broadening determined by small-angle scattering can be used to explain the measured proton beam profiles. It should be noted that although the model seems to be able to calculate the broadening of the forward-propagating fast-electron distribution gener\-ated by intense laser--matter interaction, it could fail to determine the real number of electrons arriving at the rear side of the target. According to Davies \cite{Davies2002}, the generation of electromagnetic fields, as well as the recirculation of the electrons, must be taken into account, making estimation and even calculation very difficult. Recent experiments by Akli \textit{et al.} \cite{akli2007} have shown that this is true at least for targets thinner than \(20\Uum\), but for thicker foils the assumption of strong recirculation overestimates the number of electrons. Therefore, the question of whether electromagnetic fields and recirculation are essential to determine fast-electron transport from the front to the rear side of the target can still not be satisfactorily answered, although assuming simple collisional broadening gives a a relatively good estimate.

\subsection{Target normal sheath acceleration}
\label{sec:Target Normal Sheath Acceleration - TNSA}

The electrons are transported through the target to its rear side. The laser creates about \(10^{13}\) electrons, which potentially all propagate through the target. The broadening results in transverse extension, which can be estimated by
\begin{equation}
r_{\textrm{sheath}} = r_0 + d \tan(\theta /2),
\end{equation}
where \(r_0\) denotes the laser spot radius, \(d\) the target thickness, and \(\theta\) the broadening angle of the distri\-bution, \eg calculated using \Eref{eq:broaden}. The electrons exhibit an exponential energy distribution
\begin{equation}
n_{\textrm{hot}}(E)=n_0 \exp \left( - \frac{E}{k_\textrm{B}T_{\textrm{hot}}} \right)
\end{equation}
with temperature \(k_\textrm{B} T\) and overall density \(n_0\), as given by \Eref{eq:total_el}. The electron density at the rear side (neglecting recirculation), therefore, can be estimated as

\begin{align}
n_{\textrm{e},0}& = \frac{\eta E_\textrm{L}}{c\tau_\textrm{L} \pi (r_0 +d \tan\theta/2)^2 k_\textrm{B}T_{\textrm{hot}}}\label{eq:edens1}\\
&\approx 1.5\times 10^{19} \frac{r_0^2}{(r_0+d \tan \theta /2)^2}\frac{I_{18}^{7/4}}{\sqrt{1+0.73I_{18}\lambda_{\UumZ}^2}-1}~\left[\Ucm^{-3}\right]~.
\label{eq:edens}
\end{align}
This last approximation was obtained by inserting \(E_0 = \sqrt{2I_0/\epsilon_0 c}\approx 2.7 \times 10^{12}\UV/\UmZ \) and using Eqs. (\ref{eq:total_el}) and (\ref{eq:frac}), together with a practical notation for the electron temperature based on ponderomotive scaling,
\begin{equation} k_\textrm{B}T_{\textrm{hot}}= m_0c^2 \left(  \sqrt{1+\frac{I_0 ~\left[\UWZ/\UcmZ^2\right]~\lambda_\textrm{L}^2~\left[\UumZ^2\right]}{1.37 \times 10^{18}}}-1 \right)~,
\label{eq:pond}
\end{equation}
in \Eref{eq:edens1}. \(I_{18}\) indicates that the intensity has to be taken in units of \(10^{18}\UW/\UcmZ^2\). The estimate shows that the electron density at the rear side of the target strongly scales with the laser intensity and is inversely proportional to the square of the target thickness. Taking the standard example of a laser pulse with \(I = 10^{19}\UW/\UcmZ^2\), focused to a spot of \(r_0 = 10\Uum\) and assuming a target thickness \(d = 20\Uum\), the angular broadening according to \Eref{eq:broaden} is \(\theta = 42^\circ\) (FWHM) for electrons with mean energy \(k_\textrm{B} T\), as determined by \Eref{eq:pond}. Hence, the electron density at the target's rear side is \(n_{\textrm{e},0} = 1.4 \times 10^{20}\UcmZ^{-3}\). This is orders of magnitude below the density of the solid and justifies the assumption of a shielded transport through the target.

The electrons arrive at the rear side of the target and escape into the vacuum. The charge separ\-ation leads to an electric potential \(\Phi\) in the vacuum region, according to Poisson's equation. In one dimension, it is given as
\begin{equation}
\epsilon_0 \frac{\partial^2 \Phi}{\partial z^2}=en_\textrm{e}~.
\label{poisson}
\end{equation}
To solve \Eref{poisson}, it is assumed that the solid matter in one half-space \((z \leq 0)\) perfectly compensates for the electric potential, whereas for \(z \rightarrow \infty\) the potential goes to infinity. The derivative \(\partial \Phi/\partial z\) vanishes for \(z \rightarrow \pm \infty\).
In the vacuum region \((z > 0)\), the field can be obtained analytically \cite{crow1975}. The electron density is taken as
\begin{equation}
n_\textrm{e} = n_{\textrm{e},0}  \exp \left( \frac{e\Phi}{k_\textrm{B}T_{\textrm{hot}}} \right)~,
\end{equation}
where the electron kinetic energy is replaced by the potential energy \(-e\Phi\). The initial electron dens\-ity \(n_{\textrm{e},0}\) is taken from \Eref{eq:edens}. The solution of the Poisson equation is found by using the ansatz \(e\Phi/k_\textrm{B} T_{\textrm{hot}} = -2 \ln(\lambda z + 1)\), where \(\lambda\) is a constant defined by the solution and the `\(+1\)' is necessary to fulfil a continuous solution with the condition \(\Phi(0) = 0\) at the boundary to the solid matter. The resulting potential is
\begin{equation}
\Phi(z)=-\frac{2k_\textrm{B}T_{\textrm{hot}}}{e} \ln \left( 1+ \frac{z}{\sqrt2 \lambda_\textrm{D}} \right)
\end{equation}
and the corresponding electric field reads
\begin{equation}
E(z)=\frac{2k_\textrm{B}T_{\textrm{hot}}}{e}\frac{1}{z+\sqrt2 \lambda_\textrm{D}}~.
\label{eq:field}
\end{equation}

In this solution, the electron Debye length
\begin{equation}
\lambda_\textrm{D}=\left( \frac{\epsilon_0k_\textrm{B}T_{\textrm{hot}}}{e^2n_{\textrm{e},0}} \right)^{1/2}
\end{equation}
appears; this is defined as the distance over which significant charge separation occurs \cite{krall1986}. Replacing \(k_\textrm{B} T_{\textrm{hot}}\) with \Eref{eq:pond} and \(n_{\textrm{e},0}\) with \Eref{eq:edens} leads to
\begin{equation}
\lambda_\textrm{D} \approx 1.37\Uum \frac{r_0+d \tan \theta/2}{r_0}\frac{\sqrt{1+0.73I_{18}\lambda^2}-1}{I_{18}^{7/8}}~.
\label{Debyelength}
\end{equation}

The Debye length, or longitudinal sheath extension, on the rear side is of the order of a micrometre. It scales quadratically with target thickness (since \(d \tan(\theta /2) \propto d^2\)) and is inversely proportional to the laser intensity. Thus, a higher laser intensity on the front side of the target leads to a shorter Debye length at the rear side and results in a stronger electric field. The standard example of \Eref{Debyelength} leads to \(\lambda_\textrm{D} = 0.6\Uum\).

The maximum electric field is obtained at \(z = 0\):\\
\begin{align}
E_{\textrm{max}}(z=0)&=\frac{\sqrt2k_\textrm{B}T_{\textrm{hot}}}{e\lambda_\textrm{D}}\\
&\approx 5.2 \times 10^{11}\UV/\UmZ~ \frac{r_0}{r_0 + d \tan \theta /2} I_{18}^{7/8}~,\\
&= 9 \times 10^{10}\UV/\UmZ~\frac{r_0}{r_0+d \tan \theta /2} E_{12} E_{12}^{3/4}.
\label{eq:maxelectricfield}
\end{align}
Hence, the initial field at \(z = 0\) is proportional to the laser intensity and depends almost quadratically on the laser's electric field strength. In \Eref{eq:maxelectricfield}, the laser's electric field strength is inserted in normalized units of \(10^{12}\UV/\UmZ\). By inserting the dependence of the broadening with target thickness from \Fref{broaden}, the scaling with the target thickness is obtained as \(E_{\textrm{max}} (z = 0) \propto d^{-2}\). The standard example leads to a maximum field strength of \(E_{\textrm{max}} \approx 2 \times 10^{12}\UV/\UmZ\) just at the surface, that is it is of the order of \(\UV[T]/\UmZ\) or \(\UV[M]/\UumZ\). It is only slightly smaller than the laser electric field strength of \(E_0 = 8.7 \times 10^{12}\UV/\UmZ\). However, at times later than \(t = 0\), the field strength is dictated by the dynamics at the rear side of the target, \eg ionization and ion acceleration.

As just mentioned, the electric field strength instantly leads to ionization of the atoms at the target's rear surface, since it is orders of magnitude above the ionization threshold of the atoms. A simple model to estimate the electric field strength necessary for ionization is the field ionization by barrier suppression (FIBS) model \cite{augst1989}. The external electric field of the laser overlaps the Coulomb potential of the atom and deforms it. As soon as the deformation is below the binding energy of the electron, the electron is instantly freed, hence the atom is ionized. The threshold electric field strength \(E_{\textrm{ion}}\) can be obtained with the binding energy \(U_{\textrm{bind}}\) as
\begin{equation}
E_{\textrm{ion}} = \frac{\pi \epsilon_0 U_{\textrm{bind}}^2}{e^3Z}~.
\end{equation}
As the electron sheath at the rear side is relatively dense, the atoms could also be ionized by collisional ionization. However, as discussed by Hegelich \cite{hegelich2002}, the cross-section for field ionization is much higher than the cross-section for collisional ionization for the electron densities and electric fields appearing at the target surface. Taking the ionization energy of a hydrogen atom with \(U_{\textrm{bind}} = 13.6\UeV\), the field strength necessary for FIBS is \(E_{\textrm{ion}} = 10^{10}\UV/\UmZ\). This  is two orders of magnitude less than the field strength developed by the electron sheath in vacuum calculated earlier. Hence, nearly all atoms (protons, carbon, heavier particles) at the rear side of the target are instantly ionized and, since they are no longer neutral particles, they are then subject to the electric field and are accelerated. The maximum charge state of ions found in an experiment is an estimate of the maximum field strength that appeared. This has been used to estimate the sheath peak electric field value \cite{hegelich2002}, as well as the transverse field extension \cite{McKenna2007, schreiber2004}.

The strong field ionizes the target and accelerates ions to mega electronvolt energies, if it is applied for a sufficiently long time. The time can easily be  calculated by the assumption of a test particle moving in a static field, generated by the electrons. Free protons were chosen as test particles. The non-linear equation of motion is obtained from \Eref{eq:field}. The solution was obtained numerically with \textsc{matlab} \cite{matlab}. It shows that for a proton to obtain a kinetic energy of 5\UMeV, the field has to stay for \(500\Ufs\) in the shape given by \Eref{eq:field}. During this time, the proton has travelled \(11.3\Uum\). The electric field will be created as soon as electrons leave the rear side of the target.

Some electrons can escape this field, whereas others with lower energy will be stopped and will be re-accelerated back into the target. Since the electron velocity is close to the speed of light and the distances are of the order of a micrometre, this happens on a time-scale of a few femtoseconds, leading to a situation where electrons are always present outside the rear side of the target. The electric field being created does not oscillate but is quasi-static on the order of the ion acceleration time. Therefore, ultra-short laser pulses, although providing the highest intensities, are not the optimum laser pulses for ion acceleration. The electric field is directed normal to the target's rear surface; hence, the direction of the ion acceleration is normal to the target, giving the process its name, target normal sheath acceleration.

\subsection{Expansion models}
\label{subsec:Expansion models}

The laser acceleration of ions from solid targets is a complicated, multidimensional mechanism including relativistic effects, non-linearities, and collective and kinetic effects. Theoretical methods for the various physical mechanisms involved in TNSA range from analytical approaches for simplified scenarios over fluid models up to fully relativistic, collisional three-dimensional computer simulations.

Most of the approaches that describe TNSA neglect the complex laser--matter interaction at the front side of the target as well as the electron transport through the foil. These plasma expansion models start with a hot-electron distribution that drives the expansion of an initially given ion distribution \cite{fuchs2006, schreiber2006, fuchs2007, mora2003, mora2005, mora20052, albright2006, crow1975, passoni2004}. Crucial features such as the maximum ion energy, as well as the particle spectrum, can be obtained analytically, whereas the dynamics have to be obtained numerically. The plasma expansion description dates back to 1954 \cite{landau1954}. Since then, various refinements of the models were obtained, with an increasing activity after the first discovery of TNSA. These calculations resemble the general features of TNSA. Nevertheless, they rely on somewhat idealized initial conditions from simple estimates. In addition to that, the plasma expansion models are one-dimensional, whereas the experiments have clearly shown that TNSA is at least two-dimensional. Hence, these models can only reproduce one-dimensional features, \eg the particle spectrum of the TNSA process.

Sophisticated three-dimensional computer simulation techniques have been developed for a better understanding of the whole process of short-pulse high-intensity laser--matter interaction, electron trans\-port and subsequent ion acceleration. The simulation methods can be classified as (i) particle-in-cell, (ii) Vlasov, (iii) Vlasov--Fokker--Planck, (iv) hybrid fluid or particle, and (v) gridless particle codes; see the short review in Ref. \cite{bell2006} for a description of each method.

The particle-in-cell method is the most widely used simulation technique. In this method,  Max\-well's equations are solved, together with a description of the particle distribution functions. The method more or less re\-sembles a `numerical experiment' with only a few approximations; hence, a detailed insight into the dynamics can be obtained. The disadvantage is that no specific theory serves as an input par\-ameter and the results must be analyzed like experimental results, i.e., they need to be interpreted and compared with analytical estimates.

\subsubsection{Plasma expansion model}
\label{subsubsec:Plasma expansion model}

Plasma expansion is often described as an isothermal rarefaction wave into free space. There is quite
a large similarity with the expansion models used to describe TNSA. The isothermal expansion model as\-sumes {\it quasi-neutrality}, \(n_\textrm{e} = Zn_\textrm{i}\), and a constant temperature \(T_\textrm{e}\). Using the two-fluid hydrodynamic model for electrons and ions, the continuity, momentum, and energy conservation equations are used, usually with the assumption of an isothermal expansion (no temperature change in time), no further source term (no laser), no heat conduction, collisions or external forces, and a pure electrostatic acceler\-ation (no magnetic fields).
One can find a self-similar solution \cite{eliezer2002}:
\begin{align}
v(z,t)& = c_\textrm{s} + \frac{z}{t}~,\\
n_\textrm{e}(z,t)&=Zn_\textrm{i}(z,t)=n_{\textrm{e},0}\exp \left( -\frac{z}{c_\textrm{s}t}-1 \right)~,
\end{align}
where \(v\) denotes the bulk velocity and \(n_\textrm{i} (n_\textrm{e})\) the evolution of the ion (electron) density. The rarefaction wave expands with the sound velocity \(c_\textrm{s}^2 = Z k_\textrm{B} T_\textrm{e}/m_\textrm{i}\).
By combining these two equations, replacing the velocity with the kinetic energy \(v^2 = 2E_{\textrm{kin}}/m\) and taking the derivative with respect to \(E_{\textrm{kin}}\), the ion energy spectrum \(\textrm{d}N/\textrm{d}E_{\textrm{kin}}\)  from the quasi-neutral solution per unit surface and per unit energy in dependence of the expansion time \(t\) is obtained \cite{mora2003}:
\begin{equation}
\frac{\textrm{d}N}{\textrm{d}E_{\textrm{kin}}} = \frac{n_{\textrm{e},0}c_\textrm{s}t}{\sqrt{2Zk_\textrm{B}T_{\textrm{hot}}E_{\textrm{kin}}}} \exp \left( -\sqrt{\frac{2E_{\textrm{kin}}}{Zk_\textrm{B}T_{\textrm{hot}}}}\right)~.
\label{eq:espect}
\end{equation}
The ion number \(N\) is obtained from the ion density as \(N = n_{\textrm{e},0}c_\textrm{s}t\). Moreover, the electric field in the plasma is obtained from the electron momentum equation, \(n_\textrm{e}eE=-k_\textrm{B}T_e \nabla n_e\), as
\begin{equation}
E=\frac{k_\textrm{B}T_\textrm{e}}{ec_\textrm{s}t}= \frac{E_0}{\omega_{\textrm{pi}}t},
\label{eq:eisotherm}
\end{equation}
with \(E_0 = (n_{\textrm{e},0}k_\textrm{B}T_\textrm{e}/\epsilon_0)^{1/2}\), where \(\omega_{\textrm{pi}} = (n_{\textrm{e},0}Ze^2/m_\textrm{i}\epsilon_0)^{1/2}\) denotes the ion plasma frequency. The elec\-tric field is uniform in space (\ie constant) and decays with time as \(t^{-1}\). The temporal scaling of the velocity is obtained by solving the equation of motion, \(\dot{v} = Zq/m E\), with the electric field as before. This yields
\begin{align}
v(t)& = c_\textrm{s} \ln(\omega_{\textrm{pi}}t) + c_\textrm{s}\\
\label{eq:veloc}
z(t)& = c_\textrm{s} t \left(\ln(\omega_{\textrm{pi}} t)-1\right) + c_\textrm{s }t~.\\
\label{eq:charge}
\end{align}
Both equations satisfy the self-similar solution. The scaling of the ion density is found as \(n(t) = n_0/\omega_{\textrm{pi}} t\).

However, at \(t = 0\), the self-similar solution is not defined and has a singularity. Hence, the model of a self-similar expansion is not valid for a description of TNSA at early times and must be modified. Additionally, in TNSA there are more differences. Firstly, the expansion is not driven by an electron distribution, being in equilibrium with the ion distribution, but by the relativistic hot electrons that are able to extend in the vacuum region in front of the ions. There, quasi-neutrality is strongly violated and a strong electric field will build up, modifying the self-similar expansion solution.

Secondly, the initial condition of equal ion and electron densities must be questioned, since the hot-electron density with \(n_\textrm{e} \approx 10^{20}\Ucm^{-3}\) is about three orders of magnitude less than the solid density of the rear-side contamination layers. This argument can only be overcome by the assumption of a global quasi-neutrality condition \(Zn_\textrm{i} = n_\textrm{e}\).

Thirdly, it might not be reasonable to assume a model of an isothermal plasma expansion. It can be assumed, however, that the expansion is isothermal, since the laser pulse provides `fresh' electrons from the front side of the target, \ie the assumption is valid for the laser pulse duration \(\tau_\textrm{L}\). As will be shown, the main acceleration time period is of the order of the laser pulse duration. This justifies the assumption of an isothermal expansion.

The plasma expansion, including charge separation, was quantitatively described by Mora \cite{mora2003,mora2005, mora20052} with high accuracy. The main point of this model is a plasma expansion with charge separation at the ion front, in contrast with a conventional, self-similar plasma expansion. The plasma consists of electrons and protons, with a step-like initial ion distribution and an electron ensemble that is in thermal equilibrium with its potential. The megaelectronvolt electron temperature results in a charge separation being present for long times. It leads to enhanced ion acceleration at the front, compared with the case of a normal plasma expansion. This difference is sometimes named the TNSA effect.

Although only one-dimensional, the model has been successfully applied to experimental data at more than 10 high-intensity short-pulse laser systems worldwide in a recent study \cite{fuchs2006}. It was separately used to explain measurements taken at the ATLAS-10 at the Max-Planck Institute in Garching, Ger\-many \cite{kaluza2004} as well as to explain results obtained at the VULCAN PW \cite{robson2007} (with a few modifications). Therefore, it is seen as a reference model, currently used worldwide for an explanation of TNSA.
Because of its success in describing TNSA, it will be explained in more detail now.

After the laser acceleration at the foil's front side, the electrons arrive at its rear side and escape into the vacuum. The atoms are assumed to be instantly field ionized, leading to \(n_\textrm{i} = n_\textrm{e}/Z\). Charge separation occurs and leads to an electric potential \(\phi\), according to Poisson's equation:
\begin{equation}
\epsilon_0 \frac{\partial^2\phi}{\partial z^2} =e \left(n_\textrm{e}(z) - n_\textrm{i}(z)\right).
\label{eq:potential}
\end{equation}
The electron density distribution is always assumed to be in local thermal equilibrium with its potential,
\begin{equation}
n_\textrm{e} = n_{\textrm{e},0} \exp \left( \frac{e \phi}{k_\textrm{B}T_{\textrm{hot}}} \right)~,
\label{eq:dens2}
\end{equation}
where the electron kinetic energy is replaced by the potential energy \(e\phi\). The initial electron density \(n_{\textrm{e},0}\) is taken from \Eref{eq:edens}. The ions are assumed be of initial constant density \(n_\textrm{i} = n_{\textrm{e},0}\), with a sudden drop to zero at the vacuum interface.
The boundary conditions are chosen such that the solid matter in one half-space \((z \leq 0)\) perfectly compensates the electric potential for \(z \rightarrow -\infty\), whereas for \(z \rightarrow \infty\) the potential goes to infinity. Its derivative \(E = -\partial\phi/ \partial z\) vanishes for \(z \rightarrow \pm \infty\).
In the vacuum region (initially \(z > 0\)), the field can be obtained analytically \cite{crow1975}. The resulting potential is
\begin{equation}
\phi(z)=-\frac{2k_\textrm{B}T_{\textrm{hot}}}{e} \ln\left( 1+ \frac{z}{\sqrt{2\exp(1)}\lambda_{\textrm{D},0}} \right) - \frac{k_\textrm{B}T_{\textrm{hot}}}{e}~,
\label{eq:analyt}
\end{equation}
and the corresponding electric field reads
\begin{equation}
E(z)=\frac{2k_\textrm{B}T_{\textrm{hot}}}{e} \frac{1}{z+ \sqrt{2\exp (1)}\lambda_{\textrm{D},0}}~.
\label{eq:analytfield}
\end{equation}

\begin{figure}
        \centering
        \includegraphics[width=0.8\textwidth]{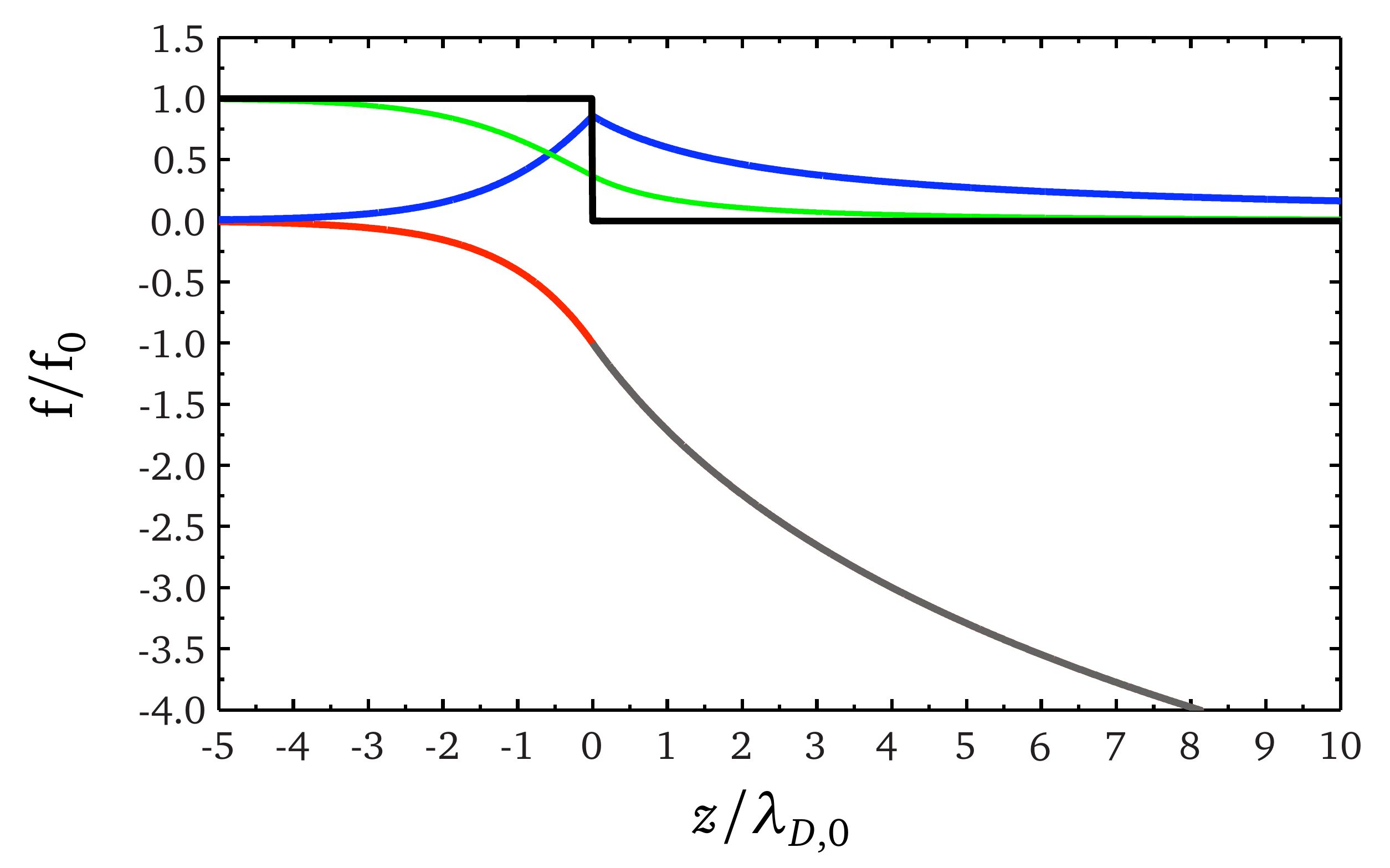}
                \caption{Solution of \Eref{eq:potential}. The potential \(\phi\) (red) was obtained numerically. The analytical solution \Eref{eq:analyt} (grey) is in perfect agreement. Both are given in units of \(k_\textrm{B} T_{\textrm{hot}}/e\). The electron density \(n_\textrm{e}\) (green), normalized to \(n_{\textrm{e},0}\), follows from \Eref{eq:dens2}. The normalized ion density \(n_\textrm{i}\) (black) is a step function with \(n_\textrm{i}(z < 0)/n_{\textrm{e},0} = 1\) and zero for \(z > 0\). The electric field E (blue) is given in units of \(k_\textrm{B}T_{\textrm{hot}}/e\lambda_{\textrm{D},0}\). The \(z\) coordinate is given in units of \(\lambda_{\textrm{D},0}\).}
\label{potential}
\end{figure}

The initial electron Debye length is \(\lambda_{\textrm{D},0}^2 = \epsilon_0k_\textrm{B} T_{\textrm{hot}}/e^2n_{\textrm{e},0}\). The full boundary value problem, including the ion distribution, can only be solved numerically. The result obtained with \textsc{matlab} \cite{matlab} is shown in \Fref{potential}. The potential \(\phi\) (red curve) is a smooth function and is in perfect agreement with the analytical solution \Eref{eq:analyt} (grey curve) in the vacuum region. Both are given in units of \(k_\textrm{B} T_{\textrm{hot}}/e\). The electron density \(n_\textrm{e}\) (green curve), normalized to \(n_{\textrm{e},0}\), follows from \Eref{eq:dens2}. The normalized ion density \(n_\textrm{i}\) (in black) is a step function with \(n_\textrm{i}(z < 0)/n_{\textrm{e},0} = 1\) and zero for \(z > 0\). The electric field \(E\) (blue curve) has a strong peak at the ion front, with \(E_{\textrm{max}} = \sqrt{2/ \exp(1)} E_0 = 0.86 E_0\) . The normalization field \(E_0\) is given by \(E_0 = k_\textrm{B} T_{\textrm{hot}}/e\lambda_{\textrm{D},0}\). The \(z\) coordinate was normalized with \(\lambda_{\textrm{D},0}\).
The subsequent plasma expansion into vacuum is described in the framework of a fluid model, governed by the equation of continuity (left) and the momentum balance (right):
\begin{equation}
\frac{\partial n_\textrm{i}}{\partial t}+ \frac{\partial (v_\textrm{i}n_\textrm{i})}{\partial z})=0 \hspace{2cm} \frac{\partial v_\textrm{i}}{\partial t}+v_\textrm{i}\frac{\partial v_\textrm{i}}{\partial t} =- \frac{e}{m_\textrm{p}} \frac{\partial \phi}{\partial z}~.
\label{eq:fluidmod}
\end{equation}

The full expansion dynamics can only be obtained numerically. Of particular interest is the tem\-poral evolution of the ion distribution and the evolution of the electric field driving the expansion of the bulk. In \cite{Schollmeierdiss} a Lagrangian code in \textsc{matlab} was developed, which solves Eqs.~(\ref{eq:dens2}), (\ref{eq:analyt}), and (\ref{eq:fluidmod}), similar to Ref. \cite{mora2003}. The numerical method is similar to the method described in Ref. \cite{kaluza2004}; however, the developed code uses \textsc{matlab}'s built-in bvp4c function for a numerical solution of the boundary value problem in the ion fluid. The initially constant ion distribution is divided into a grid, choosing the left boundary to be \(L \gg c_\textrm{s} t\). The boundary value for the potential is \(\phi(-L) = 0\). At the right boundary (initially at \(z = 0\)), the electric field \(-\partial \phi_{\textrm{front}}/\partial z = \sqrt{2/e} k_{\textrm{B}}T_{\textrm{hot}}/e\lambda_{\textrm{D,front}}\)  must coincide with the analytical solution of \Eref{eq:analytfield}, where the local Debye length must be determined by the potential at the front:
\begin{equation}
\lambda_{\textrm{D,front}}=\lambda_{\textrm{D},0} \exp \left( \frac{e\phi_{\textrm{front}}}{k_\textrm{B}T_{\textrm{hot}}} \right)^{-1/2}~.
\end{equation}

Initially, the Debye length at the ion front is obtained by inserting \Eref{eq:analyt} in \Eref{eq:dens2} to give \(\lambda_{\textrm{D,0,front}} = e^{-1}\lambda_{\textrm{D},0}\).
The code divides the fluid region into a regular grid. Each grid element (cell) has position \(z_j\), ion
density \(n_j\), and velocity \(v_j\). For each time-step \(\Delta t\), the individual grid elements are moved according to the following scheme \cite{kaluza2004}:
\begin{align}
z_{j'}& =z_j+v_j\Delta t+ \frac{e}{2m_\textrm{p}} E \Delta t^2~,\\
v_{j'}& = v_j + \frac{e}{m_\textrm{p}} E \Delta t~.
\end{align}
After that, the density of the cell is changed according to the broadening of the cell due to the movement:
\begin{equation}
n_{j'} =n_j \frac{\Delta x_j}{\Delta x_{j'}}~.
\end{equation}
At the front, the individual cells quickly move forward, resulting in a `blow-up' of the cells, which dramatically diminishes the resolution. Thus, after each time-step, the calculation grid is mapped onto a new grid ranging from \(z_{\textrm{min}}\) to the ion front position \(z_{\textrm{front}}\) with an adapted cell spacing. This method is called rezoning. The new values of \(v_j\) and \(n_j\) are obtained by third-order spline interpolation, providing very good accuracy.

\subsubsection{Temporal evolution and scaling}
\label{subsubsec:Temporal evolution and scaling}

A crucial point in the ion expansion is the evolution of the electric field strength \(E_{\textrm{front}}\), the ion velocity \(v_{\textrm{front}}\) and the position \(z_{\textrm{front}}\) of the ion front. Expressions given by Mora are \cite{mora2003,mora2005, mora20052}\\
\begin{align}
E_{\textrm{front}} &\simeq \left( \frac{2n_{\textrm{e},0}k_BT_{\textrm{hot}}}{\textrm{e} \epsilon_0} \frac{1}{1+\tau^2} \right)^{1/2}~,
\label{eq:efront}\\
v_{\textrm{front}}& \simeq 2c_\textrm{s} \ln \left( \tau + \sqrt{1+\tau^2} \right),
\label{eq:vfront}\\
z_{\textrm{front}}& \simeq  2 \sqrt{2\textrm{e} } \lambda_{\textrm{D},0} \left[ \tau \ln \left( \tau + \sqrt{1+\tau^2} \right) - \sqrt{1+\tau^2} +1 \right],
\label{eq:zfront}
\end{align}
where \(\textrm{e} = \exp(1)\) and \(\tau = \omega_{\textrm{pi}} t/\sqrt{2\textrm{e}}\). The other variables in these equations are the initial ion density \(n_{\textrm{i},0}\), the ion-acoustic (or sound) velocity \(c_\textrm{s} =(Zk_\textrm{B}T_{\textrm{hot}}/m_\textrm{i})^{1/2}\), \(T_{\textrm{hot}}\) is the hot-electron temperature and \(\omega_{\textrm{pi}} =(n_{\textrm{e},0}Ze^2/m_\textrm{i} \epsilon_0)^{1/2}\) denotes the ion plasma frequency. Owing to the charge separation, the ion front expands more than twice as quickly as the quasi-neutral solution in Eqs.~(\ref{eq:veloc}) and (\ref{eq:charge}).
From \Eref{eq:vfront}, the maximum ion energy is given as\\
\begin{equation}
E_{\textrm{max}} =2k_\textrm{B}T_{\textrm{hot}} \ln^2 \left( \tau+\sqrt{1+\tau^2} \right).
\label{eq:emax}
\end{equation}
The particle spectrum from Mora's model \cite{mora2003,mora2005, mora20052} cannot be given in an analytic form, but it is very close to the spectrum of \Eref{eq:espect}, obtained by the self-similar motion of a fully quasi-neutral plasma expanding into a vacuum. The phrase `fully quasi-neutral' should indicated that in this solution there is no charge separation at the ion front, hence there is no peak electric field.

A drawback of the model is the infinitely increasing energy and velocity of the ions with time, which results from the assumption of an isothermal expansion. Hence, a stopping condition must be defined. An obvious time duration for the stopping condition is the laser pulse duration \(\tau_\textrm{L}\). However, as found by Fuchs \textit{et al.} \cite{fuchs2006, fuchs2007}, the model can be successfully applied to measured maximum energies and spectra, as well as to particle-in-cell simulations, if the calculation is stopped at \(\tau_{\textrm{acc}} = \alpha(\tau_\textrm{L} + t_{\textrm{min}})\). It was found that, for very short pulse durations. the acceleration time \(\tau_{\textrm{acc}}\) tends towards a constant value \(t_{\textrm{min}} = 60\Ufs\), which is the minimum time that the energy
transfer from the electrons to the ions needs. The variable \(\alpha\) takes into account that for lower laser intensities the expansion is slower and the acceleration time must be increased. \(\alpha\)\ varies linearly from 3 at an intensity of \(I_\textrm{L} = 2 \times 10^{18}\UW/\UcmZ^2\) to \(1.3\) at \(I_\textrm{~L} = 3 \times 10^{19}\UW/\UcmZ^2\). For higher intensities, \(\alpha\) is constant, at 1.3. Hence, the acceleration time is
\begin{equation}
\tau_{\textrm{acc}}= \left( -6.07 \times 10^{-20} \times (I_\textrm{L} - 2 \times 10^{18})+3\right) \times (\tau_\textrm{L} + t_{\textrm{min}})
\end{equation}
for \(I_\textrm{L} \in [2 \times 10^{18} - 3 \times 10^{19}]\UW/\UcmZ^2\),
and\begin{equation}
\tau_{\textrm{acc}} = 1.3 \times(\tau_\textrm{L} + t_{\textrm{min}})
\end{equation}
 for \(I_\textrm{L} \geq 3 \times 10^{19}\UW/\UcmZ^2\).

\begin{figure}
        \centering
        \includegraphics[width=0.8\textwidth]{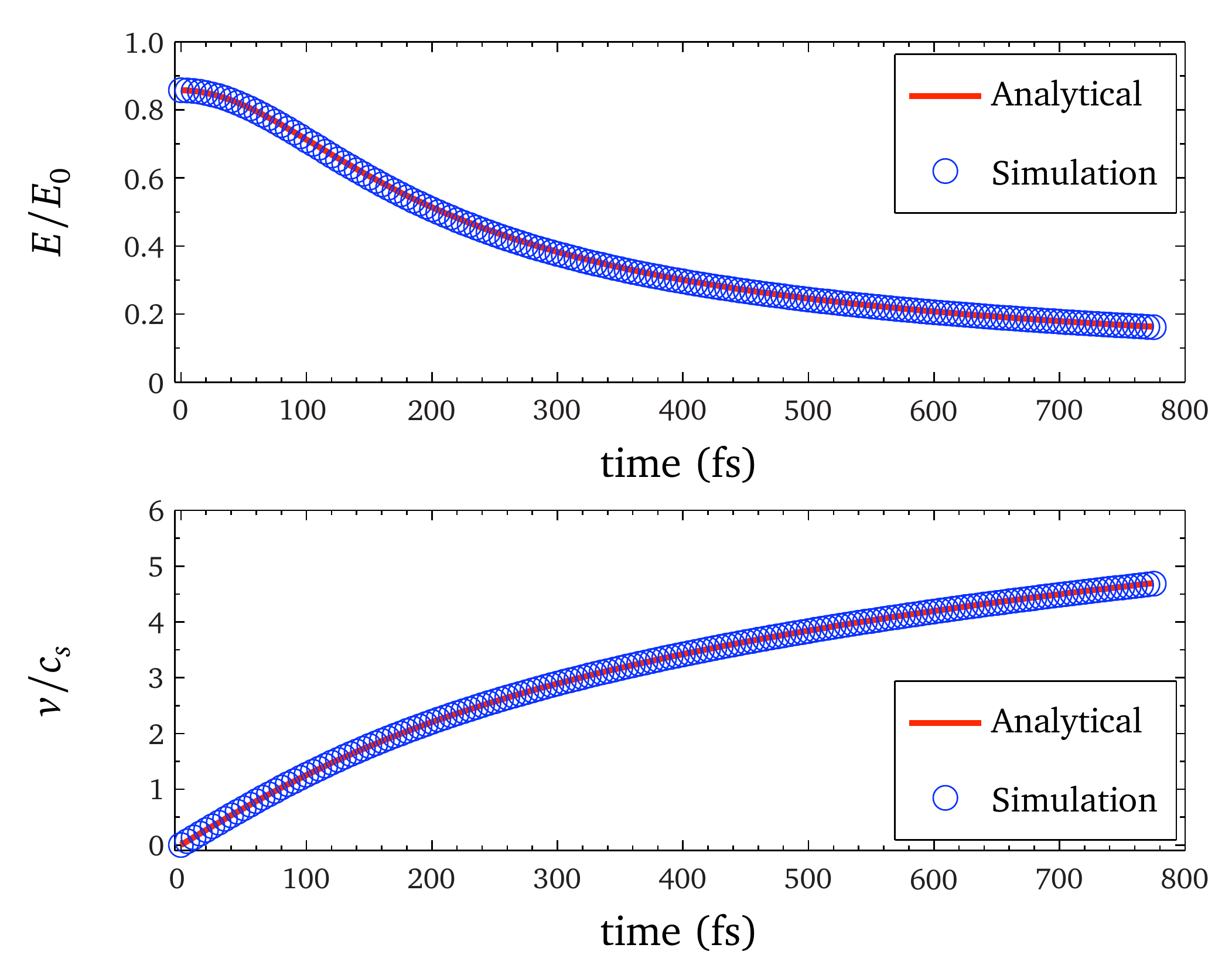}
                \caption{Temporal evolution of the electric field and the ion velocity at the ion front. There is very good agreement between simulated values (blue circles) and Eqs.~(\ref{eq:efront}) and (\ref{eq:vfront}) (red curve).}
\label{tempevosim}

\end{figure}

\begin{figure}
        \centering
        \includegraphics[width=1.0\textwidth]{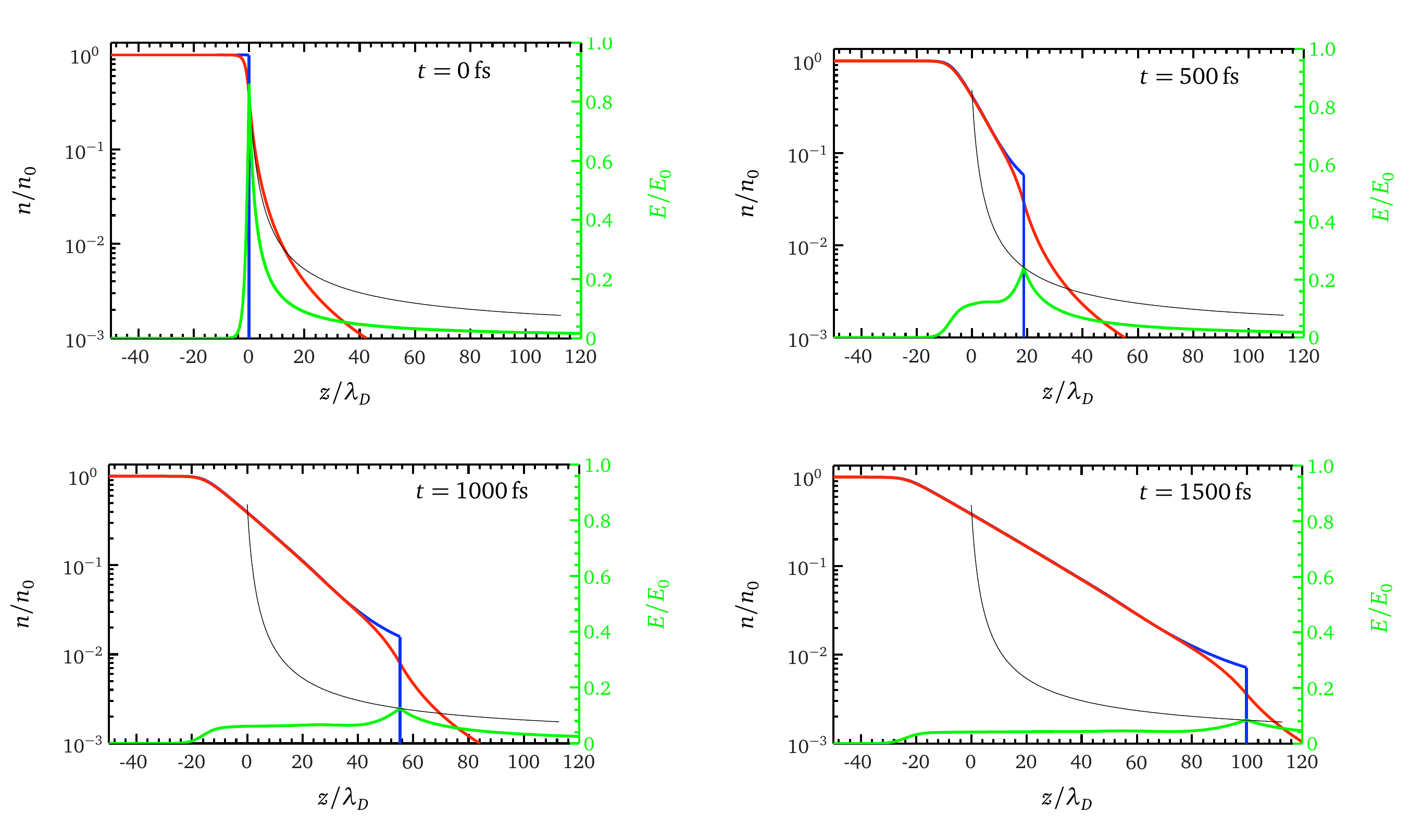}
                \caption{Temporal evolution of the electric field and the ion and electron density, respectively. The electric field (green curve) peaks sharply at the ion front. The ion density (blue curve) is \(n_\textrm{i} =n_0\) for \(z\leq 0\) and zero for \(z > 0 \) for \(t = 0\). The electron density (red curve) decays proportionally to \(z^{-2}\). For later times, at \(t = (500, 1000, 1500)\Ufs\), the ions are expanded, forming an exponentially decaying profile.}
\label{tempelfsim}
\end{figure}

\Figure[b]~\ref{tempevosim} shows the temporal evolution of the electric field and the ion velocity at the ion front, respectively. The electric field was normalized to \(E_0\), the ion velocity is divided by the sound velocity. There is a very good agreement between the simulated values (blue circles) and the expressions by Mora \cite{mora2003,mora2005, mora20052} from Eqs.~(\ref{eq:efront}) and (\ref{eq:vfront}) (red curve). The maximum deviations from the scaling expressions are 1.6\% for the electric field and 0.4\% for the velocity.
The electric field evolution and the development of the electron and ion density profiles are shown in \Fref{tempelfsim}. The electric field (green curve) peaks sharply at the ion front for all times. Initially, the ion density (blue curve) is \(n_\textrm{i} =n_0 \) for \(z\leq 0\) and zero for \(z>0\). The electron density (red curve) is infinite and decays proportionally to \(z^{-2}\). Note the different axis scalings for the electric field and the densities; the latter are plotted on a logarithmic scale. For later times, at \(t = (500, 1000, 1500)\Ufs\), the ions are expanded, forming an exponentially decaying profile.

A large part of the expanding plasma is quasi-neutral and can be identified by the constant electric field, as derived in \Eref{eq:eisotherm}. At the ion front, the charge separation is still present, leading to an enhanced electric field that is a factor of two greater than the electric field in the bulk, in agreement with Ref. \cite{mora2003}. This scaling is maintained for the whole expansion. The scaling of the peak electric field value at the ion front at position \(z\), as given by the analytical expressions in Eqs.~(\ref{eq:efront}) and (\ref{eq:zfront}), is in perfect agreement with the simulation.

\begin{figure}
        \centering
        \includegraphics[width=0.75\textwidth]{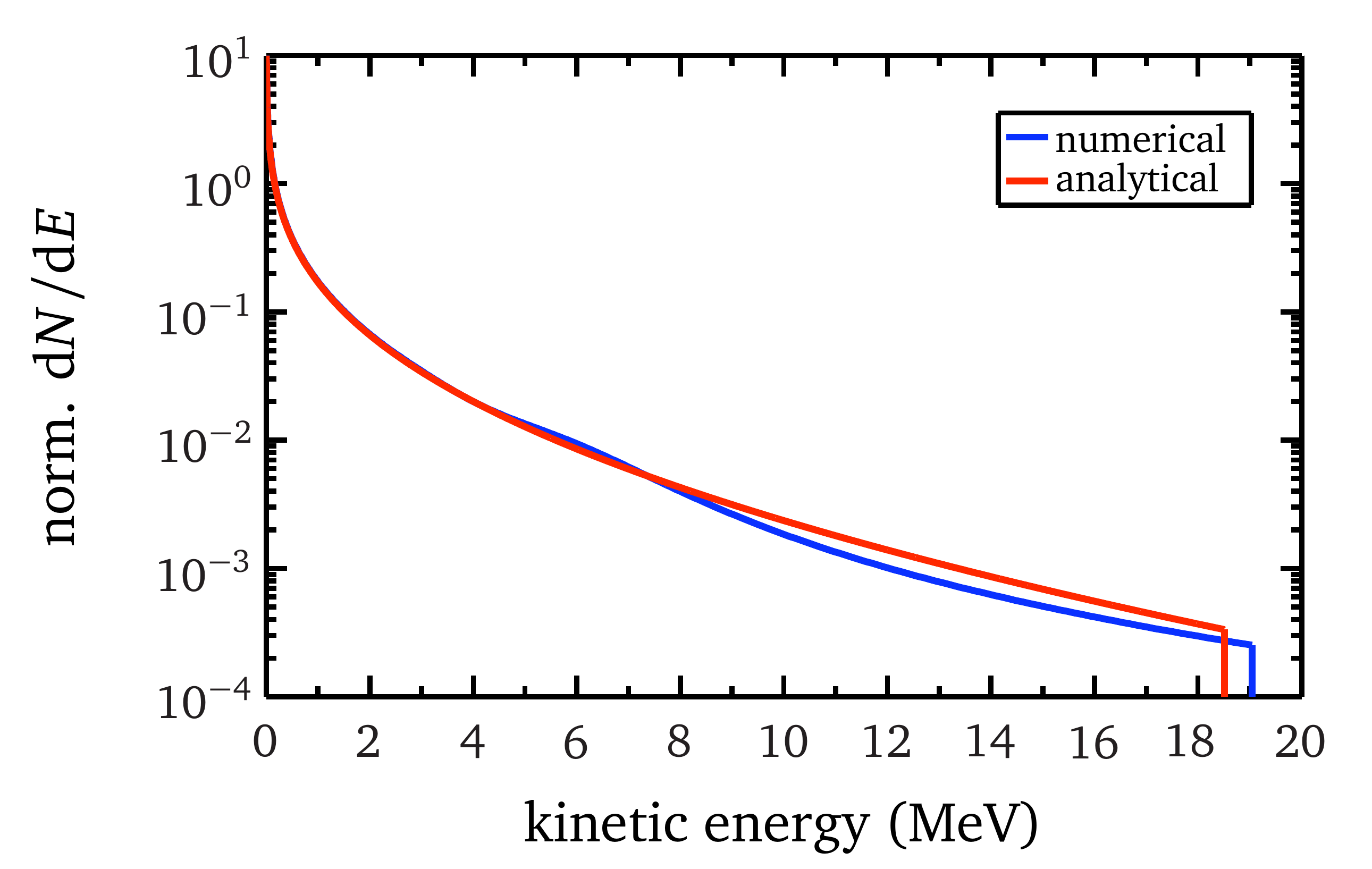}
                \caption{Energy spectrum \(\textrm{d}N/\textrm{d}E\) from the simulation (blue curve) compared to the spectrum of a quasi-neutral plasma expansion (red curve). norm, normalized.}
\label{specsim}
\end{figure}

The final proton spectrum is shown in \Fref{specsim}. The numerical solution (blue curve) is close to the analytical solution from the quasi-neutral model of \Eref{eq:espect} (red curve). The analytical spectrum is assumed to reach a maximum energy, taken from \Eref{eq:emax}. The maximum energy in the simulation is \(E_{\textrm{max,num}} = 19\UMeV\), which is in close agreement to the analytical value of \(E_{\textrm{max,analyt}} = 18.5\UMeV\). As expected, there is excellent agreement in the spectra for low energies, since in both cases the expansion is quasi-neutral. For high energies, the numerical spectrum deviates from the self-similar model. The numerical spectrum is lower than the self-similar one even though the ion density of the numerical solution increases close to the ion front, as can be seen in \Fref{tempelfsim} in the deviation of the electron and ion densities close to the front. However, the velocity increase at the front in the simulation is much faster than that in the self-similar solution, owing to the enhanced electric field. Thus, the kinetic energy of the fluid elements close to the ion front is greater than the kinetic energy of fluid elements in a self-similar expansion. The spectrum is obtained by taking the derivative of the ion density with respect to the kinetic energy. It turns out that the kinetic energy increases more strongly than the ion density, hence \(\textrm{d}N/\textrm{d}E\) is a little less than the self-similar expansion.

In conclusion, the Lagrangian code and the model developed by Mora show that TNSA-accelerated ions are mainly emitted in the form of a quasi-neutral plasma, with a charge separation at the ion front that leads to an enhanced acceleration compared with the expansion of a completely quasi-neutral plasma. For later times, if \(\omega_{\textrm{pi}} t \gg 1\), the analytical expression of the maximum ion energy in \Eref{eq:emax} can be used to determine the cut-off energy of TNSA-accelerated ions accurately. The spectral shape of the ions is close to the spectrum of a quasi-neutral, self-similar expansion.

The equations show, that the maximum energy, as well as the spectral shape, strongly scale with the hot-electron temperature. The expression for the initial electric field scales as \(E \propto k_\textrm{B}T_{\textrm{hot}}n_\textrm{e}\); hence, a simplistic estimate would assume that both are equally important for the maximum ion energy. In contradiction, however, the investigation has shown that the maximum ion energy only depends weakly on the hot-electron density and is directly proportional to the hot-electron temperature. It is worth noting that this finding is in agreement with results obtained earlier with an electrostatic particle-in-cell code by Brambrink \cite{brambrink2004}.
The hot-electron density---owing to the quasi-neutrality boundary condition---determines the number of generated ions. Both the number of ions and the energy increase with time, again showing that not the shortest and most intense laser pulses are favourable for TNSA, but some\-what longer pulses, of the order of a picosecond. This requires a high laser energy to keep the intensity sufficiently high.

Nevertheless, the model is still very idealized, since it is one-dimensional and isothermal, with the electrons ranging into infinity, and it neglects laser interaction and electron transport. An approach with electrons in a Maxwellian distribution always leads to the same asymptotic behaviour of the ion density \cite{gurevich1966}, hence two-temperature \cite{passoni20042} or even tailored \cite{kumar2008} electron distributions will lead to different ion distributions. There are many alternative approaches to the one described here, including, \eg an adiabatic expansion \cite{mora20052}; multitemperature effects \cite{mora20052, passoni20042}; an approach in which an upper integration range is introduced to satisfy the energy conservation for the range of a test electron in the potential \cite{passoni2004}; expansion of an initially Gaussian-shaped plasma \cite{mora2005}; and expansion of a plasma with an initial density gradient \cite{grismayer2006}. Most of these approaches assume an underlying fluid model, where particle collisions are neglected and the fluid elements are not allowed to overtake each other. Hence a possible wave\-breaking or accumulation of particles is not included in the models but requires a kinetic description  \cite{grismayer2008, bychenkov2004}. Furthermore, the transverse distribution of the accelerated ions cannot be determined from a one-dimensional model and requires further modelling. This can be done in the framework of  two-dimensional particle-in-cell simulations. Particle-in-cell simulations allow a much more sophisticated description, including relativistic laser--plasma interaction, a kinetic treatment of the particles, and a fully three-dimensional approach.

\section{Target normal sheath acceleration: ion beam characteristics}
\label{sec:TNSA - Ion beam characeristics}

Part of the motivation of the extensive research on laser-accelerated ion beams is based on their ex\-ceptional properties (high brightness and high spectral  cut-off, high directionality and laminarity, short pulse duration), which distinguish them from those of the lower-energy ions accelerated in earlier ex\-peri\-ments at moderate laser intensities. In view of these properties, laser-driven ion beams can be employed in a number of groundbreaking applications in science, technology, and medicine. This section reviews the main beam parameters; Section \ref{sec:Applications} focuses on established and proposed applications using these unique beam properties.

\subsection{Beam parameters}
\label{subsec:Beam Parameter}

\subsubsection{Particle numbers}
One of the striking features of TNSA-accelerated ion beams is the fact that the particle number in a forward-directed beam is very high. At present, particle numbers of up to \(6 \times 10^{13} \) protons with energies above \(4\UMeV\) have been detected in experiments. This typically leads to a conversion efficiency of laser to ion beam energy of up to 9\%. At these high particle numbers, drawn from a very limited source size, for high-energy short-pulse laser systems, the depletion of the proton contamination layer at the rear surface becomes an issue. This has been addressed by Allen \textit{et al.} \cite{allen2004}, who determined that there are \(2.24 \times 10^{23}\)~atoms\(/\UcmZ^3\) at the rear surface of a gold foil, in a layer 1.2~\text{\AA} thick. Assuming an area of about \(200\Uum\) diameter, the accelerated volume is about \(V = 3.8 \times 10^{-11}\Ucm^{-3}\). Hence, the total number of protons in this area is about \(N_{\textrm{total}} = 8.4 \times 10^{12}\), which is close to the integrated number determined in the experiments. Experiments have shown \cite{roth2002} that a rear surface coating of a metal target can provide enough protons up to a thickness of \(\approx\)100\Unm, where the layer thickness causes the onset of instabilities in the electron propagation, owing to its limited electrical conductivity.

\subsubsection{Energy spectrum}
Based on the acceleration mechanism and the expansion model described earlier, the usual ion energy distribution is an exponential one with a cut-off energy that is dependent on the driving electron tempera\-ture. Without special target treatment, and independently of the target material, protons are always acceler\-ated first, as they have the highest charge-to-mass ratio. These protons stem from water vapour and hydro\-carbon contamination, which are always present on the target surface, owing to the limited achievable vacuum conditions. Protons from the top-most contamination layer on the target surface are exposed to the highest field gradients and screen the electric field for protons and ions coming from the successive layers. The acceleration of particles from different target depths results in a broad energy distribution, which becomes broader with increasing contamination layer thickness. The inhomogeneous electron distribution in the sheath also leads to an inhomogeneous accelerating field in the transverse direction. The resulting exponential ion energy spectrum constitutes the main disadvantage in laser ion acceleration.

Only three groups so far \cite{hegelich2006, schwoerer2006, ter-avetisyan2006} have produced a quasi-monoenergetic ion beam with lasers and an energy spread of 20\% or less. Hegelich \textit{et al.} \cite{hegelich2006} have used 20\(\Uum\) thick palladium foils that were resistively heated before the acceleration. At temperatures of about \(600\UK\), the targets were completely dehydrogenized, but carbon atoms still remained on the surface. By increasing the target temperature \((T > 1100\UK)\), the carbon underwent a phase transition and formed a monolayer or graphite (graphene) on the palladium surface, from which \(C^{5+}\) ions were accelerated to \(3\UMeV/\textrm{u}\) with an energy spread of 17\%. An advantage of resistive heating is the complete removal of all hydrogen at once but there are also several disadvantages. The formation of graphene cannot be controlled and the set-up requires a very precise temperature measurement.

Schwoerer \textit{et al.} \cite{schwoerer2006} used \(5\Uum\) thick titanium foils coated with \(0.5\Uum\) thick hydrogen-rich poly\-methylmethacrylate dots of \(20\Uum\times 20\Uum\) on the target's rear surface. This configuration was designed to limit the transverse extension of protons, so that the proton-rich dot would have a smaller diameter than the scale of inhomogeneity of the electron sheath. In this case, all protons experience the same potential. The parasitic proton contamination layer could be reduced by nanosecond-laser ablation; the accelerated protons showed a quasi-monoenergetic energy spectrum peaking at \(1.2\UMeV\).

Ter-Avetisyan \textit{et al.} \cite{ter-avetisyan2006} produced quasi-monoenergetic deuteron bursts by the interaction of a high-intensity, high-contrast (\({>}10^{-8}\)) laser with limited-mass water droplets. The peak position in the spectrum, at 2\UMeV, had an energy spread of 20\%. This experiment, however, suffered from a low laser--droplet interaction probability.

\subsubsection{Opening angle}
\label{Openingangle}
\begin{figure}
        \centering
        \includegraphics[width=0.75\textwidth]{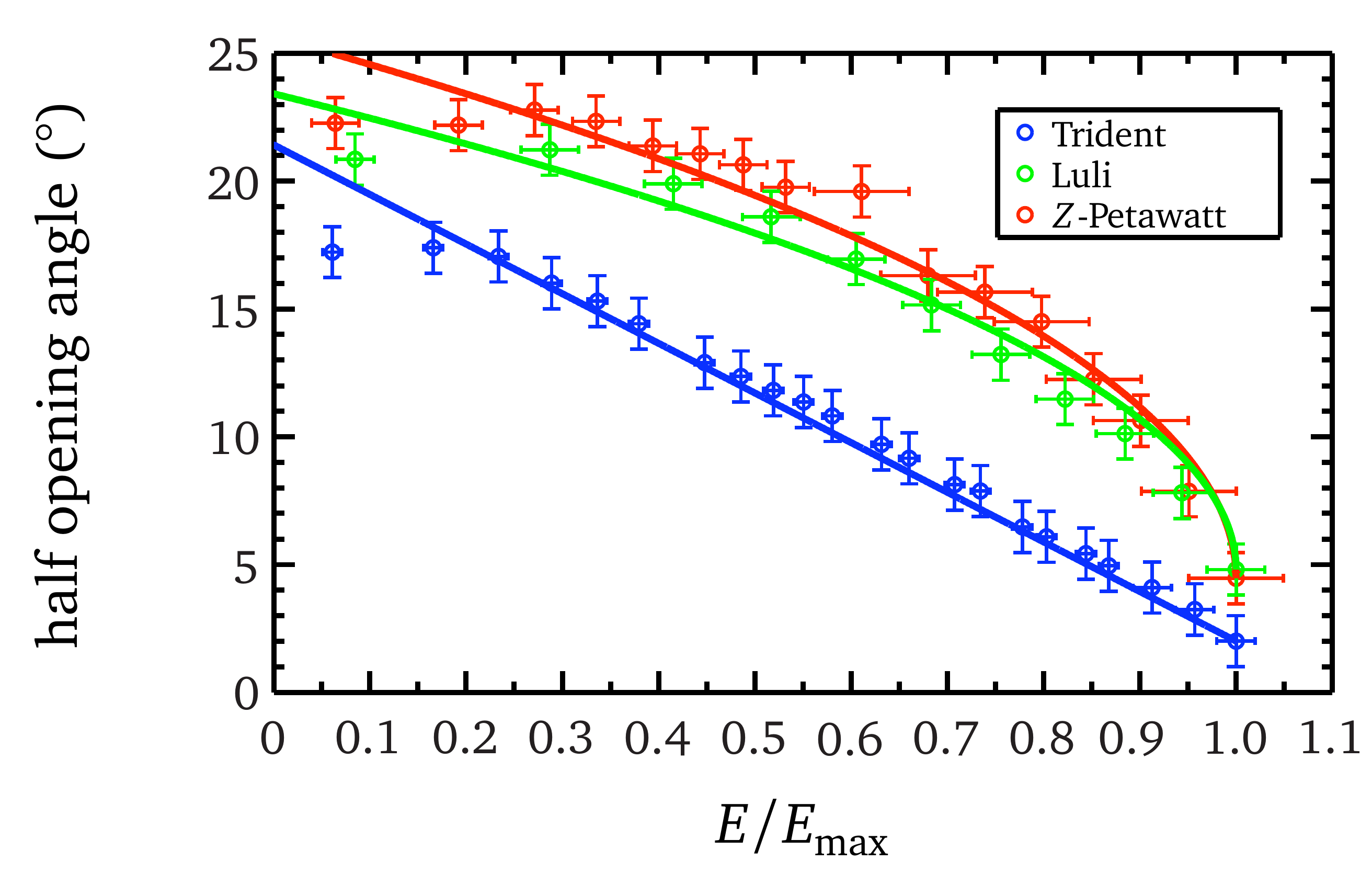}
                \caption{Energy dependence of the half opening angle. Data obtained at Trident (blue circles), LULI-100\UW[T] (green circles), and Z-Petawatt (red circles). The plots have been normalized to the respective maximum energy of each beam. The opening angle decreases with increasing energy. A parabolic dependency could be fit to the LULI and Z-Petawatt results, the data for Trident produce a linear slope.}
                \label{opening_angle}
\end{figure}

\Figure[b]~\ref{opening_angle} shows the energy-resolved opening angles for data obtained at Trident (blue circles), LULI-100\UW[T] (green circles), and Z-Petawatt (red circles). The plots have been normalized to the respective maximum energy of each beam. These are \(19\UMeV\) for Trident, \(16.3\UMeV\) for LULI-100\UW[T], and \(20.3\UMeV\) for Z-Petawatt, respectively.

Protons with the highest energy are emitted with the smallest opening angle from the source, up to a \(5^\circ\) half angle. Protons with less energy are subsequently emitted in larger opening angles. Below about 30\% maximum energy, the opening angle reaches a maximum and stays constant for lower energies. In most cases, the opening angles decrease parabolically with increasing energy, as shown in \Fref{opening_angle}. In some shots, however, the decrease of opening angle with increasing energy is close to linear, as in the example obtained at Trident. The slope of the opening angle with energy is a result of the initial hot-electron sheath shape at the target surface, as pointed out by Carroll \textit{et al.} \cite{carroll2007}. According to Ref. \cite{carroll2007}, a sheath with Gaussian dependence in the transverse direction results in a strongly curved opening angle--energy distribution, whereas a parabolic hot-electron sheath results in a linear dependency. However, only crude details of the exact modelling of the acceleration process are given in Ref. \cite{carroll2007}.
In Ref. \cite{Schollmeierdiss}, a more detailed expansion model is described, which is able to explain the experimental results in more detail.
It should be noted that the term `opening angle' is not equivalent to the beam `divergence'. The divergence of the protons increases slightly with increasing energy, whereas the emitting area (source size) decreases with proton energy \cite{borghesi2006, cowan2004}. This results in a total decrease of the opening angle measured experimentally.

\subsubsection{Source size}
\begin{figure}
        \centering
        \includegraphics[width=0.75\textwidth]{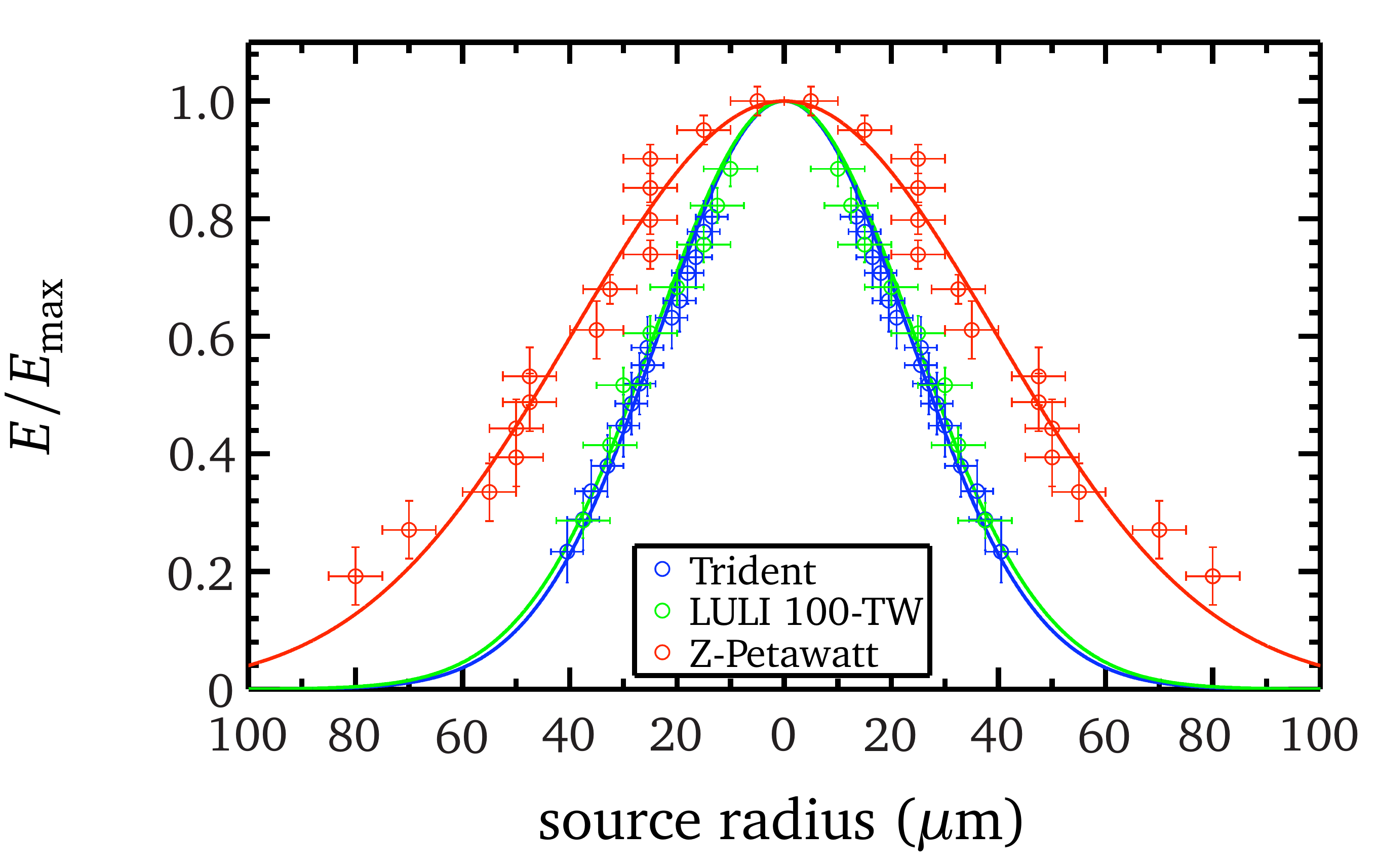}
                \caption{Energy-resolved source sizes for data from Trident (blue circles), LULI-100\UW[T] (green circles) and Z-Petawatt (red circles). The energy-source-size distribution fits to a Lorentzian with  full width at half maximum  of \(54.8\Uum\) for Trident, \(56.5\Uum\) for LULI-100\UW[T], and \(92.8\Uum\) for Z-Petawatt.}
                \label{source_size}
\end{figure}

\Figure[b]~\ref{source_size} shows energy-resolved source sizes for the three laser systems Trident (blue circle), LULI-100\UW[T] (green circles), and Z-Petawatt (red circles). As in Section \ref{Openingangle}, the energy axis has been normalized to the individual maximum energy of the shot, with the maximum energies as before. Source size decreases with increasing energy. Protons with the highest energies are emitted from sources of about \(10\Uum\) diameter and less. For lower energies, the source sizes progressively increase, up to about \(200\Uum\) diameter for the lowest energies measurable with radiochromic film imaging spectroscopy, about \(1.5\UMeV\). For even lower energies, the source sizes might be much larger and could reach in excess of \(0.5\Umm\) in diameter \cite{schreiber2004}.

             The energy dependence of the source size well fits a Gaussian, indicated in \Fref{source_size}. The data could be fit by
\begin{equation}
E = \exp \left( \frac{-(4\ln(2) {\mbox{source size}})^2}{\sigma^2} \right)~,
\end{equation}
where \(2 \sigma\) denotes the FWHM. This fit allows the complete energy-dependent source size to be char\-acter\-ized using only one parameter. The FWHM for Trident with a \(10\Uum\) thin gold target is \(\sigma = 54.8\Uum\). For LULI-100\UW[T], the source size is \(\sigma = 56.5\Uum\) for a \(15\Uum\) thin gold foil. A larger source size has been measured at Z-Petawatt, with \(\sigma = 92.8\Uum\) and a \(25\Uum\) thick gold target.

The energy dependence of the source size is directly related to the electric field strength distri\-bution of the accelerating hot-electron sheath at the source. Protons with high energies have been acceler\-ated by a high electric field. Cowan \textit{et al.} \cite{cowan2004} relate increasing source size with decreasing energy to the shape of the hot-electron sheath, under the assumption of an isothermal quasi-neutral plasma expansion where the electric field is \(E = -(k_\textrm{B}T_\textrm{e}/e)(\delta n_\textrm{e}/n_\textrm{e})\). A transversely Gaussian electric field distribution would result in a non-analytical expression for the electron density, \(n_\textrm{e}\). Conversely, a realistic assumption of a Gaussian hot-electron distribution would result in a radially linearly increasing electric field, in contradiction to the measured data. Hence, it is concluded that the quasi-neutral plasma expansion, although it is the driving acceleration mechanism for later times, does not explain the observed source sizes.
In fact, the source size must develop earlier in the acceleration process, \eg at very early times, when the electric field is governed by the Poisson equation (\Eref{poisson}), with \(E(z) \propto k_\textrm{B}T_\textrm{e}/\lambda_\textrm{D} \propto \sqrt{k_\textrm{B}T_\textrm{e}n_\textrm{e}} \). With the data from \Fref{source_size}, it can be concluded that there must be a radial dependency of \(E(z)\), and hence a Gaussian decay of either the hot-electron temperature or the density, or both.

\subsubsection{Emittance}
As we have seen, the acceleration of the ions by the TNSA mechanism basically constitutes a quasi-electrostatic field acceleration of initially cold (room-temperature)
atoms at rest, which are field ionized and then pulled by the charge-separation field. As the electrons are scattered while being pushed through the target, at least for materials with enough electrical conductivity to provide the return currents, the transport smooths the distribution at the rear surface to result in a Gaussian-like field shape that expands laterally in time as the electrons start to recirculate. Thus the initial random ion movement in the beam, represented in an extended phase space and often regarded as an effective beam temperature, is very low.

An important parameter in accelerator physics is the transverse emittance of an ion beam. In view of the nature of the ion sources used in conventional accelerators, there is always a spread in kinetic energy and velocity in a particle beam. Each point on the surface of the source emits protons with different initial magnitude and direction of the velocity vector. The emittance \(\epsilon\) provides a figure of merit for describing the quality of the beam, \ie its laminarity \cite{humphries2008}.
Assume that the beam propagates in the \(z\)-direction. Each proton represents a point in the position-momentum space \((x, p_x\) and \(y, p_y )\), the phase space. The transverse phase space (\eg in the \(x\)-direction) of the TNSA protons is obtained by mapping the source position (indicated for example by imprinted surface grooves in the radiochromic
film) versus the angle of emission \(p_x/p_z\), obtained by the position \(x\) of the imprinted line in the radiochromic
film and the distance \(d\) by \(p_x/p_z =x' =\arctan(x/d)\).

In general, the quality of charged-particle beams is characterized by their emittance, which is proportional to the volume of the bounding ellipsoid of the distribution of particles in phase space. By Liouville's theorem, the phase  space volume of a particle ensemble is conserved during non-dissipative acceleration and focusing. For the transverse phase space dimensions, the area of the bounding phase space ellipse equals \(\pi \epsilon\), where the emittance \(\epsilon\), at a specific beam energy (or momentum \(p\)), is expressed in normalized root-mean-square (r.m.s.) units as
\begin{equation}
\epsilon_{\textrm{N,rms}} = (p/mc) \left[<x^2><x'^2> - <xx'>^2 \right]^{1/2}
~.
\label{emittance}
\end{equation}

In \Eref{emittance}, \(m\) is the ion mass, \(c\) is the velocity of light, \(x\) is the particle position within the beam envelope, and \(x' = p_x/p_z\) is the particle's divergence in the \(x\)-direction. At a beam waist, \Eref{emittance} reduces to \(\epsilon_{\textrm{N,rms}} = \beta \gamma \sigma_x \sigma_{x'}\), where \(\sigma_x\) and \(\sigma_{x'}\) are the r.m.s.~values of the beam width and divergence angle. Several effects contribute to the overall emittance of a beam: its intrinsic transverse `thermal' spread; intrabeam space charge forces; and non-ideal accelerating fields, for example, at apertures in the source or accelerator. For typical proton accelerators (\eg the CERN SPS or FNAL-Tevatron), the emittance at the proton source is \({\approx}0.5\Umm\cdot\UmradZ\) (normalized r.m.s.) and up to 20--80\(\Umm\cdot\Umrad\) within the synchrotron. The longitudinal phase space \((z\)--\(p_z)\) is characterized by the equivalent energy--time product of the beam envelope;  for the CERN SpS, a typical value  is \({\sim}0.1\UeV\cdot\UsZ\).

The highest quality ion beams have the lowest values of transverse and longitudinal emittance, indicating a low effective ion temperature and a high degree of angle-space and time-energy correlation. In typical TNSA experiments, one may estimate an upper limit for the transverse emittance of the proton beam from \Eref{emittance}, by assuming that initial beamlets were initially focused to a size \({\ll}100\UnmZ\), and that the observed width is entirely due to the initial width of the \(x'\) distribution. The upper limit of the emittance for \(12\UMeV\) protons is \({<}0.002\Umm\cdot\UmradZ\). This is a factor of \({>}100\) smaller than typical proton beam sources, and we can attribute this to the fact that during much of the acceleration the proton space charge is neutralized by the co-moving hot electrons, and that the sheath electric field self-consistently evolves with the ions to produce an effectively `ideal' accelerating structure. The remaining irreducible `thermal' emittance would imply a proton source temperature of \({<}100\UeV\). The energy spread of the laser-accelerated proton beam is large, ranging from zero to tens of megaelectronvolts; however, owing to the extremely short duration of the accelerating field (\({<}10\Ups\)),\, the longitudinal phase space energy--time product must be less than \(10^{-4}\UeV\cdot\UsZ\). More details can be found in Ref. \cite{cowan2004}, from which part of this section was extracted.

\subsubsection{Ion species}
Since protons are the lightest ions, and have the highest charge-to-mass ratio, they are favoured by the acceleration processes. The protons are present in the target as surface contaminants or in compounds of the target itself or of the target coating. The cloud of accelerated protons then screens the electric field generated by the electrons for all the other ion species. The key for the efficient acceleration of heavy ions is the removal of any proton or light ion
contaminants. In a few cases, heating of the target was performed prior to the experiments, to eliminate the hydrogen contaminants as much as possible, and to obtain a better, more controllable ion acceleration. In particular, by removing the proton from the targets, or by choosing H-free targets, the acceleration of heavier ions was favoured. For the latter case, recent experiments have demonstrated heavy-ion acceleration of up to more than \(5\UMeV/\)u, which corresponds to ion energies that are usually available at the end of a conventional accelerator of hundreds of metres in length \cite{hegelich2002}.

\begin{figure}
        \centering
        \includegraphics[width=0.9\textwidth]{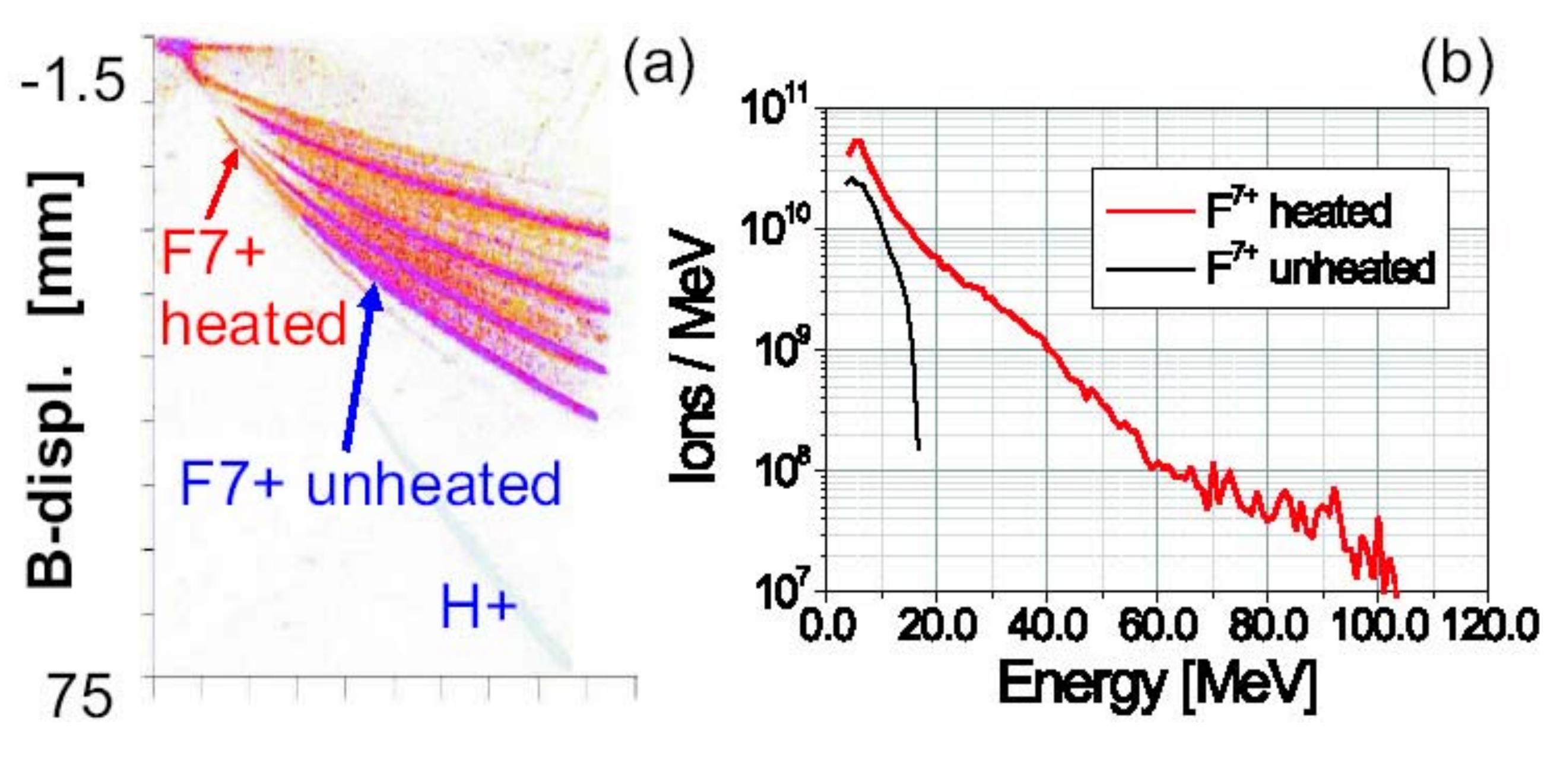}
                \caption{(a) Overlaid signals of heated and non-heated (blue) \(\EW/\ECa\EF_2\) targets: the proton signals vanish for heated targets, the fluorine signals (especially \(\EF^{7+}\)) go up to much higher energies. (b) Corresponding \(\EF^{7+}\) spectra: more than \(5\UMeV\) per nucleon are achieved for \(\EF^{7+}\) ions. From Ref. \cite{hegelich2002}.}
                \label{heavy_ion}
\end{figure}
First attempts to remove the hydrogen contaminants used resistively heated Al targets to tempera\-tures of a few hundred degrees Celsius. The partial removal of the hydrogenous contaminants already strongly enhanced the acceleration of carbon ions \cite{roth2002}. A reduction by a factor of ten increased the energy of the carbon ions by a factor of \(2.5\) and their number by two orders of magnitude. Using tungsten as a thermally stable target resist and coating the rear surface with the material of interest, the target could be heated to more than 1200 degrees Celsius. Such targets show no accelerated protons at all, but instead a strongly increased contribution of heavy ions. The maximum energy could be enhanced by a factor of five compared with aluminium targets and the conversion efficiency again by a factor of ten \cite{hegelich2002} (see \Fref{heavy_ion}). Where the ohmic heating of target materials of interest is prohibited, owing to a low evaporation point, laser heating has proved an appropriate option for removing the contamination layers. In that case, the intensity of the laser heating the rear surface and evaporating the proton layers and the timing, with respect to the short pulse, have to be matched carefully. The energy spectrum of the heavy ions, together with their charge-state distribution also provides detailed information about the accelerating electric field at the rear surface. It was shown that, in a typical experiment, collisional ionization and recombination in flight is
negligible, and so the detected charge states directly image the electric field strength because of the field ionization process. The field strengths that have been obtained match the estimated field strength, also predicted by theory, very well, and range from \(10^{11}\UV/\UmZ\) to \({\sim}10^{12}\UV/\UmZ\).

The accelerating field deduced from ion acceleration is also consistent with observed proton ener\-gies with non-heated targets. For example, in typical experiments, fluorine ions were accelerated up to \(100\UMeV\), \ie more than \(5\UMeV\)/nucleon at a maximum charge state of \(7^+\). This corresponds to an electric field of \(2\UV[T]/\UmZ\), which would have accelerated protons, if present, to energies of up to \(25\UMeV\). These were exactly the maximum energies found in experiments with non-heated targets under similar experimental conditions. Furthermore, the conversion efficiency is very high. Similar to the results obtained for proton beams, conversion efficiencies of up to 4\% from laser to ion beam energy have been measured. Because of the same accelerating mechanism for protons and heavy ions, an excellent beam quality was expected. This section was extracted from \cite{borghesi2006}.

\subsection{Target dependence}
\label{subsec:Target dependence}

In the previous section, we have looked extensively at the influence of target thickness and conductivity on the driving electron sheath distribution. To summarize, a highly conducting ultra-thin target is the most favourable for efficient ion acceleration. Moreover, as the electrons can distribute a part of the energy provided by the laser into bremsstrahlung, a low-\(Z\) material is preferable. The ultimate thickness of the target is determined by the limited laser contrast, as TNSA requires a sharp density gradient at the rear surface. For effective ion acceleration, an undisturbed back surface of the target is crucial to provide a sharp ion density gradient, since the accelerating field strength is proportional to \(T_\textrm{hot}/el_0\), where \(T_\textrm{hot}\) is the temperature of the hot electrons and \(l_0\) is the larger of either the hot-electron Debye length or the ion scale length of the plasma on the rear surface. The limited contrast of the laser causes a shockwave, launched by the prepulse, that penetrates the target and causes a rarefaction wave that diminishes the density gradient at the back, thereby drastically reducing the accelerating field. The inward moving shockwave also alters the initial conditions of the target material due to shockwave heating and therefore changes, \eg the target density and conductivity. As a trade-off, one has, however, to note that a certain preplasma at the front side is beneficial to the production of hot electrons, somehow contradicting the need for high-contrast lasers. Also, as the lateral expansion of the electron sheath affects the evolution of the ion acceleration, it has been found that to confine the electron by reducing the transverse dimensions of the targets also enhances the ion particle energy. So the ideal target would resemble an ultra-thin, low-\(Z\), highly conducting target with small lateral dimensions and a large preplasma at the front side.

Meanwhile, high-contrast laser systems are able to irradiate targets as thin as only a few nano\-metres, and we begin to see the change in the accelerating mechanism
to radiation pressure acceleration \cite{robinson2008} or laser breakout afterburner type acceleration \cite{yin2007}, which is beyond the scope of this paper.

Apart from maximizing the ion beam particle energy, targets can be used to shape the beam for the applications listed next. Ballistic focusing \cite{roth2002, patel2003, dyer2008, antici2006, snavely2007} and defocusing \cite{roth2002} have been demonstrated by numerous groups, tailoring on a nanoscale using microstructured targets \cite{brambrink2006} and layered targets to modify the shape of the energy spectrum\cite{schwoerer2006}.

\subsection{Beam control}
\label{subsec:Beam Control}

Ballistic focusing of laser-accelerated proton beams has been known since 2003 \cite{patel2003} and has been in\-vesti\-gated in detail because of the large importance for proton-driven fast ignition \cite{roth2001} and the generation of warm dense matter \cite{pelka2010}. Experiments have shown that the real source size of a few hundred micro\-metres could be collimated down to \(30\Uum\) using ballistic focusing from the rear surface of hemispherical targets. However, one has to take into account the real sheath geometry of the driving electrons to inter\-pret the proton beam profile accurately. The driving sheath consists of a superposition of the sheath field normal to the inner surface of the hemisphere and the Gaussian electron density distribution caused by the limited source size of the driving laser focus. Therefore, the experiments in Ref. \cite{patel2003} have indicated that a larger laser focal spot should minimize the second effect and thus result in a better focus quality.

For almost all applications, a precise control of the beam parameter and the possibility of tailoring the beams is of great importance. As we have seen, using the guidance by the shape of the target material, we have large control over the spatial ion beam distribution. However, it might also be instructive and preferable to try to manipulate the ion beam just using optical methods.
Results reported so far were obtained with a round laser spot, focused as well as possible to obtain the highest intensities. But, as found by Fuchs \textit{et al.}~\cite{fuchs2003}, the laser focal spot shape is eventually imprinted in the accelerated proton beam. Fuchs \textit{et al.}~assumed that the bulk of the hot electrons follow the laser focal spot topology and create a sheath with the same topology at the rear side. The proton beam spatial profile, as detected by a film detector, was simulated with a simple electrostatic model. Fuchs \textit{et al.}~took the laser beam profile as an input parameter and assumed the electron transport to be homogeneous, with a characteristic opening angle that needed to be fit to match the measured data. The unknown source size of the protons was fit to best match the experimental results. It was shown, that for their specific target thickness and laser parameters, the fitted broadening angle of the electron sheath at the rear side closely matched the broadening angle expected by multiple Coulomb small-angle scattering. However, Fuchs \textit{et al.}~could only fit the most intense part of the measured beam and neglected the lower intense part that originates from rear-side accelerated protons. Additionally, there is no information on the dependence of these findings on target thickness.

Using microgrooved targets, a more detailed understanding has been achieved. The asymmetric laser beam results in asymmetric proton beam profiles. The energy-resolved source size of the protons was deduced by imaging the beam perturbations from the microgrooved surface into a radiochromic
film detector stack. It was shown that the protons with the highest energies were emitted from the smallest source. When the laser focus size was increased, the proton source size increased as well. For symmetric as well as asymmetric laser beam profiles, the source-size dependent energy distribution in both cases could be fit to a Gaussian. This leads to the conclusion that the laser beam profile has no significant contribution to the general expansion characteristics of laser-accelerated protons, but can strongly modify the transverse beam profile without changing the angle of the beam spread. For a more detailed analysis of the experimental results, a code for  sheath-accelerated beam ray-tracing for ion analysis (SABRINA) was developed, which takes the laser beam parameters as input and calculates the shape of the proton distribution in the detector. The electron transport was modelled to follow the laser beam profile topology closely and a broadening due to small-angle collisions was assumed. It was shown that broadening due to small-angle collisions is the major effect that describes the source size of protons for thick target foils \((50\Uum)\). By contrast, thin target foils \((13\Uum)\) show much larger sources than expected for small-angle collisions. The physical reason behind this observation remains unclear; it is most likely the result of electron refluxing.
Thus, the shape of the sheath at the rear side of the thick targets can be estimated by a simple model of broadening due to multiple small-angle scattering, but the model fails for the description of sheath broadening in thin targets.

The imprint of the laser beam profile affects the intense part of the proton beam profile. This effect must also be present in cases with a round focal spot. Therefore, a focal spot with a sharp peaked laser beam profile will result in a strongly divergent proton beam, as observed in experiments. The findings also explain that in cases where a collimation of the proton beam is required, \eg proton fast ignition or the injection of the beam into a post-accelerator, not only is a curved target surface necessary, but a large, flat-top laser focal spot is indispensable, to produce a flat proton-accelerating sheath.

\section{Applications}
\label{sec:Applications}

Summarizing the beam parameters achievable using the TNSA mechanism one can make a number of conclusions.

The measured particle energies so far extend up to tens of megaelectronvolts (\(78\UMeV\) protons, \(5\UMeV\)/u palladium) and the particles showed complete space charge and current neutralization due to accompanying electrons. In experiments, particle numbers of more than \(10^{13}\) ions per pulse and beam currents in the mega-amp regime were observed. Another outstanding feature is the excellent beam quality, with a transverse emittance of less than \(0.004 \pi \Umm\cdot\Umrad\) and a longitudinal emittance of less than \(10^{-4}\UeV\cdot\UsZ\). Because of these unmatched beam characteristics, a wealth of applications were foreseen immediately. Those applications range from:
\begin{itemize}
\item injection of high power ion beams for large-scale basic research facilities;
                                               \item new diagnostic techniques for short-pulse phenomena, since the short pulse duration allows for the imaging of transient phenomena;
\item modification of material parameters (from applications in material science to warm dense matter research and laboratory astrophysics);
\item applications in energy research (`fast igniter' in the inertial fusion energy context);
\item medical applications;
\item use as a new laser-driven pathway to compact, bright neutron sources.
\end{itemize}

\subsection{Ion source}
\label{subsec:Ion Source}

An important application for TNSA ion beams is for use as next-generation ion sources or accelerators, where the excellent beam quality and strong field gradients can replace more conventional systems. Several collaborations are actively working on that task, spearheaded by the LIBRA collaboration in the UK, the LIGHT collaboration in Germany and a group at JAEA in Japan.

We briefly address a few aspects of current research aimed at applying laser ion beams as a new source.

\subsubsection{Collimation and bunch rotation of accelerated protons}
One of the main drawbacks of laser-accelerated ions and, in particular, protons are the exponential energy spectrum and the large envelope divergence of the beam.  Different techniques have been developed to modify the energy distribution, to produce a more monoenergetic beam. Therefore, special targets were created with thin proton or carbon layers on the rear side, as well as deuterated droplets.  In addition, there have been attempts to reduce the initial divergence of the beams by ballistic focusing, with the help of curved hemispherical targets, resulting in a beam focus at a distance of the diameter of the sphere from the laser focus. In a different experiment, a laser-triggered microlens was used to select a small energy interval and to focus the protons with these specific energies to a millimetre spot \(70\Ucm\) from the target \cite{toncian2006}.

\begin{figure}
        \centering
        \includegraphics[width=0.8\textwidth]{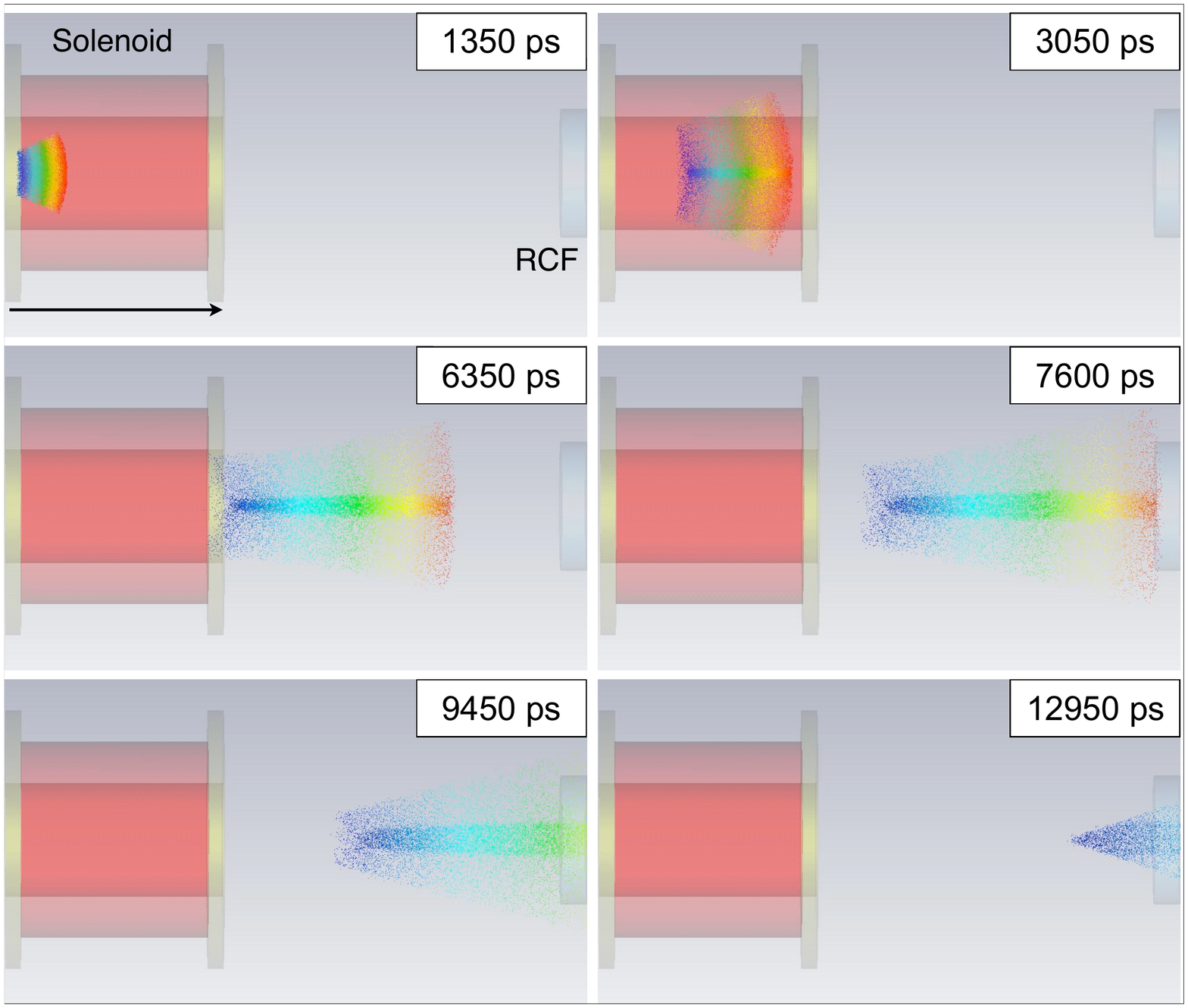}
                \caption{Simulation of the propagation of target normal sheath acceleration protons through a solenoid magnet up to a detector at six timepoints. For a better view, the accompanying electrons are not shown. A clear aggregation of protons at the axis due to space charge effects is visible. The proton energy in this case ranges from 1 to 5\UMeV.  RCF, radiochromic film. Published in Ref. \cite{harres2010}.}
                \label{solenoid}
\end{figure}

The total proton yield of typically \(10^{13}\) particles and the extremely high observed phase space dens\-ity immediately behind the source and prior to any collimator are highly encouraging. As in all cases of sources of secondary particles (antiprotons, muons, rare isotopes, \etc), transmission efficiency and phase space degradation due to the first collimator need to be carefully examined. In particular, higher than first-order focusing properties of the collimator are a serious limitation to the realistically `usable' fraction of the production energy spectrum, as well as of the production cone divergence. As these same limitations might cause a serious degradation of the transverse emittance of the `usable' protons, the very small production emittance becomes a relatively irrelevant quantity. Instead, an `effective' emittance, taking into account transmission loss and blow-up caused by the collimator, should be used. In this context, space charge (non-linear) effects are a further source of emittance degradation---probably not the dominant one---to be carefully examined.
Recently, we have shown that the collimation of a laser-accelerated proton beam by a pulsed high-field solenoid is possible and leads to good results in terms of collimation efficiencies. More than \(10^{12}\) particles were caught and transported by the solenoid. The steadiness of the proton beam after collimation could be proven up to a distance of \(324\Umm\) from the target position. Inside the solenoid, strong space charge effects occurred, owing to the co-moving electrons, which are forced to circulate around the solenoid's axis at their gyroradius by the strong magnetic field, leading to a proton beam aggregation around the axis (see \Fref{solenoid}). Details can be found in Ref. \cite{harres2010}.

More detailed calculations of the injection into ion optical structures have been published by Ingo Hofmann \cite{hofmann2009, hofmann2011}.\\

\subsubsection{Chromatic error of solenoid collimation}

In general, pulsed solenoids are a good match to the `round' production cone of laser-accelerated par\-ticles; a quadrupole-based focusing system appears to be disadvantageous in the defocusing plane of the first lens, owing to the relatively large production angles. As an example, we use the short-pulsed solenoid currently under experimental study at the GSI Helmholtz Centre for Heavy-Ion Research. It has a length of \(72\Umm\) and a theoretical maximum field strength of \(16\UT\), which is sufficient to parallelize protons at \(10\UMeV\). The distance from the target spot to the solenoid edge is assumed to be \(17\Umm\).

The prevailing higher-order effect of a solenoid is the increase in focal length with particle energy. Owing to the debunching process, different sections along the bunch have different energies and thus focus at different distances. This results in an effective increase in the bunch-averaged emittance to the effect that the tiny initial production emittance should be replaced by a chromatic emittance. To examine the expected behaviour in detailed simulation, we employed the DYNAMION code \cite{Yaramishev2006}, which includes higher-order effects in amplitudes and energy dependence, as well as space charge effects. The latter are based on particle--particle interaction, which limits the space charge resolution.
\begin{figure}
        \centering
        \includegraphics[width=0.8\textwidth]{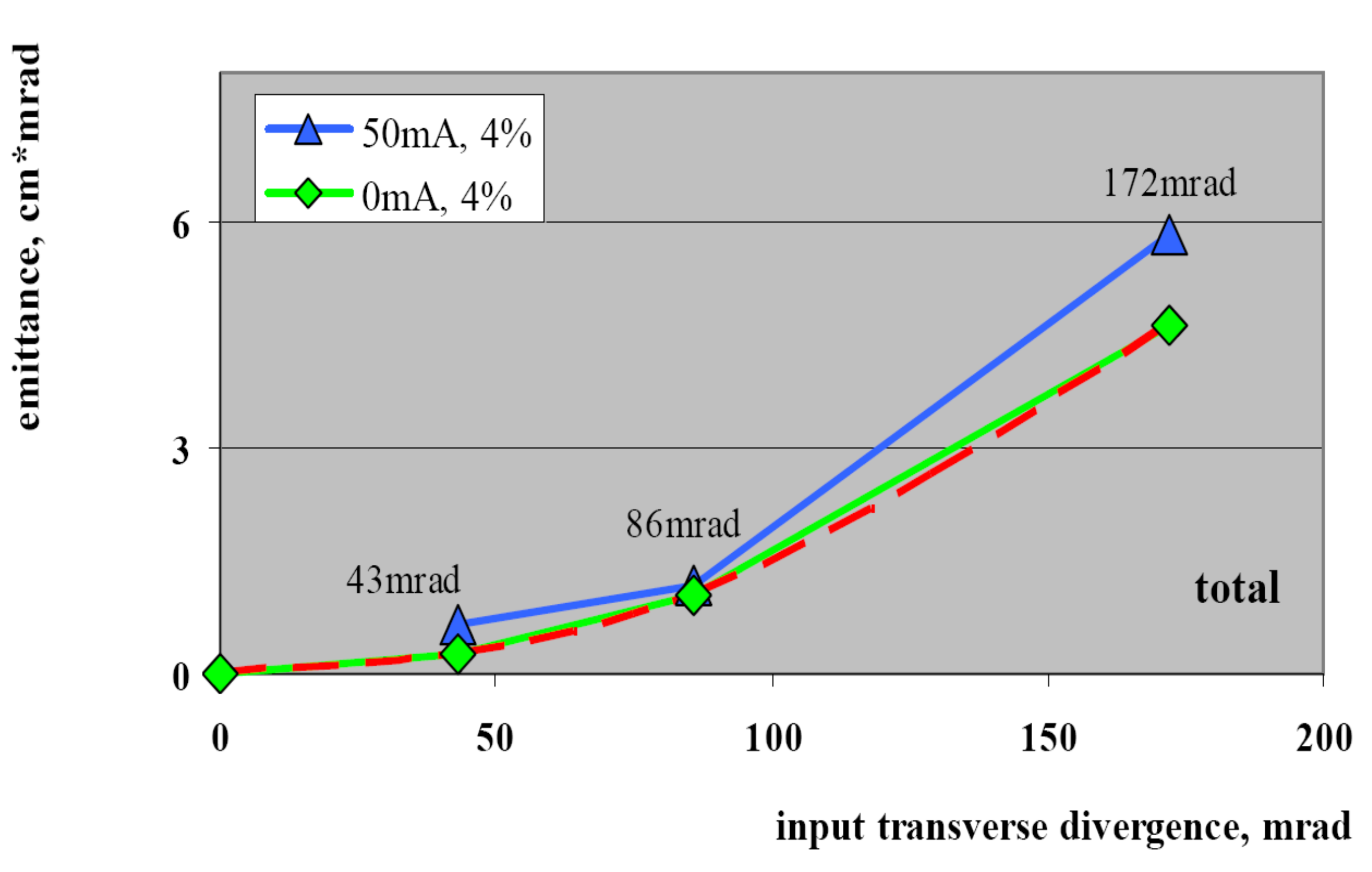}
                \caption{Dependence of `chromatic' emittances (here total emittances for 95\% of particles) on production cone angle, as obtained by DYNAMION simulations.}
                \label{chromemittance}
\end{figure}
The solenoid three-dimensional magnetic field was obtained by direct integration using the coil geometry of the ex\-peri\-mental solenoid. To quantify the chromatic effect, we consider an ensemble of protons with constant energy spread \(\Delta E/E = \pm 0.04\) around a reference energy of \(10\UMeV\). Results for final emittances (ignoring space charge) are shown in \Fref{chromemittance} to depend exactly quadratically on the considered production cone opening angle \(\delta x'\), which varied up to \( \pm 172\Umrad~( \pm 10^\circ)\). To test the influence of space charge, we also simulated a case with the number of protons in the bunch, \(N_\textrm{b}\), equal to \(3\times 10^9\), which is equivalent to a linac current of \(I=50\UmA\) (using \(I=eN_\textrm{b}f_\textrm{RF}\) and assuming that each bucket of a \(f_\textrm{RF}=108\UMHz\) sequence is filled identically). For simplicity, the bunch intensity was chosen to be independent of the opening angle. It is noted that the quadratic law still approximately applies.
Since for given \(\delta x'\) the dependence on \(\Delta E/E\) is found to be practically linear, we can justify the following scaling of the chromatic emittances in the absence of space charge:
\begin{equation}
\epsilon_x = \alpha_\textrm{c} (\delta x' )^2 \frac{\Delta E}{E},
\end{equation}
with \(\alpha_\textrm{c} \approx 0.04\Um/\UradZ\) for the particular solenoid described here. The law is still approximately true if space charge is included for the assumed bunch intensity. Note that the chromatic emittance is found to be practically independent of the initial spot radius \(r_\textrm{spot}\)---contrary to the production emittance given by the product \(r_\textrm{spot} \delta x'\). \\

\subsubsection{Transmission through beam pipe}

For planned experiments, it is important to note that the increase of emittance with energy spread will inevitably lead to transmission loss in the finite acceptance of the following beam pipe. To this end, we have assumed a beam pipe of \(3\Ucm\) radius up to \(250\Ucm\) distance from the source.  We have also assumed a linac current \(N_\textrm{b} \approx \Delta E/E\), with \(N_\textrm{b}=2 \times 10^9\) for the lowest value \(\Delta E/E = \pm 0.04\). The increase in transmission loss with distance is mostly due to the large spread of focusing angles as a function of the energy spread and is also due, to a lesser extent, to space charge. The surviving energy distribution evaluated at different distances from the source goes down to 35\% for the largest initial energy spread case in the previous example of \(\Delta E/E= \pm 0.64\) and a correspondingly high current of \(560\UmA\). Obviously, an extended beam pipe serves as an energy filter.

\subsubsection{RF bunch rotation}
For most applications of laser-accelerated particles, in particular for ion beam therapy, it is desirable to reduce the final energy spread on the target to a fraction of one per cent, to enable focusing on a small target spot. This is achieved by means of a `bunch rotation' RF cavity applied to the beam after debunching to a length suitable for the RF wavelength. The initial short bunch length increases with debunching proportional to the distance from the source and the considered energy spread. Capture into the RF bucket of a fraction of a beam within a given transverse emittance defines the ultimate six-dimensional extraction efficiency and the `usable' part of the total production of protons.  As reference value we take an energy spread of  \(\Delta E/E = \pm 0.04\), which can be reduced to a reasonably small value by using a \(500\UkV / 108\UMHz\) bunch rotation RF cavity approximately \(250\Ucm\) away from the solenoid. This means that only the central part of the totally transmitted energy distribution---about 20\% for the \(3\Ucm\) aperture limitation---can be captured by the RF bucket. Diagnosing the intensity and six-dimensional emittance of this `usable' fraction in the presence of the background of the fully transmitted spectrum is a challenge to the diagnostics.

Thus, at the current status, a careful study of the transfer efficiency of these beams into con\-ven\-tional transport and focusing structures is crucial and timely, and will be carried out within the next few years, given the unique prerequisites present among all the international collaborations. The foremost goal of the proposed effort is to determine the properties of the generated proton or ion beams with the prospect of later applications and to examine the possibilities of collimation, transport, debunching and, possibly, post-acceleration in conventional accelerator structures, both theoretically and experimentally.

\subsection{Diagnostics}
\label{subsec:Diagnostics}

A highly energetic laminar beam of charge particles with a pulse duration of only a few picoseconds constitutes an ideal tool to diagnose transient phenomena. Like a short burst of X-rays, a pulse of laser-driven protons can penetrate a target and reveal important information about its parameters. Laser-driven protons are complementary to X-rays, as the interactions of charged particles are fundamentally different from those of electromagnetic radiation. In the past, ion beams produced by conventional accelerators have already been used to radiograph static and transient samples \cite{west1972}, as well as for the in\-vesti\-gation of electric fields in laser-produced plasmas \cite{cobble2002}. Several experiments with laser-accelerated beams as probes were performed, to investigate the evolution of highly transient electric fields evolving from charging of laser-irradiated targets \cite{borghesi2003, mackinnon2004}. These fields change on a picosecond time-scale. Proton beams from ultra-intense laser--matter interactions are accelerated in a few picoseconds, depending on the laser pulse length. Combined with the very low emittance \cite{cowan2004}, these beams allow for two-dimensional mapping of the primary target with unprecedented spatial and temporal resolution. Using this technique, remnants of relativistic solitons that were generated in a laser plasma have been detected \cite{borghesi2002}, and the accelerating Debye sheath in a TNSA process as well as the ion expansion from the rear side of the target foil could be pictured \cite{borghesi2007}.

Because of the energy spectrum, and owing to the dispersion of the proton pulse at the point of interaction with the target to be probed, different proton energies probe the target at different times. Using the radiochromic
film stack technique, the ions can be separated in energy, where the high-energy particles deposit their energy in the rearmost layer, while  lower-energy particles are stopped in the front layers. Thus, by unfolding the layers, one can separate the ion energies and therefore the interaction time down to picosecond temporal resolution.

\subsubsection{Energy loss}

The fundamental contrast mechanism for generating image information is energy loss in the sample, and the consequent shift in the energy distribution of transmitted protons. As one proceeds from the shallow layers to the deepest radiochromic
film layers, protons with progressively higher incident energies are preferentially recorded.  By examining a portion of the image where the sample contained no inter\-vening material, we can deduce the incident proton energy distribution from the depth dependence of the deposited energy, based on the respective response function.

If the incident protons pass through a thin sample of thickness \(\delta x\), they lose energy according to their energy loss rate, and the transmitted proton energy, which is incident on the film detector, is

\begin{equation}
E_\textrm{f} \approx E_\textrm{i} - (\textrm{d}E/\textrm{d}x) \cdot \delta x \approx E_\textrm{i} - \Delta E~.
\end{equation}

If the sample is thick, so that \(\textrm{d}E/\textrm{d}x\) is not constant over the energy range \(\Delta E\), then the energy loss is given by the integral along the trajectory,
\begin{equation}
\Delta E = \int_0^{\delta x} \textrm{d}E(\textrm{d}x)\textrm{d}x~.
\end{equation}

A limitation of this technique is that energy loss, and therefore thickness information, is encoded as a spectral shift of the proton energy distribution due to energy loss.  If the object has a large thickness range, and hence large values of the energy loss, early-time (high incident proton energy) in\-for\-mation from thick portions of the sample is recorded in the same final proton energy interval as late-time (low incident proton energy) information from thinner portions of the sample.  Deconvolution of the spatial, temporal, and thickness information is complicated, and even self-consistent solutions may not be unique.

In the ideal limit of no transverse scattering, the resolvable thickness variation over a sample is related only to the energy loss and the exponential fall-off of the proton spectrum.  This is a strong function of initial proton energy via the energy dependence of \(\textrm{d}E/\textrm{d}x\).  For example, for our test case, in which we observe a \(2\UMeV\) exponential distribution, a resolvable intensity variation of 1\% implies a resolvable energy loss of \(20\UkeV\).  At a proton energy of \(10\UMeV\), this would correspond to a plastic (CH) thickness of \({<}5\Uum\), and, at \(4\UMeV\), a thickness of \(2\Uum\).

\subsubsection{Transverse scattering}

In addition to continuously slowing down, the protons also undergo multiple small-angle Coulomb scat\-tering from the nuclei in the sample.  In the energy range of interest, proton multiple scattering can be described by a Gaussian fit to the Moli\`{e}re distribution, which is similar to the case of the electrons we have seen earlier.  That is, protons are scattered according to a near-Gaussian polar angle distribution,
\begin{equation}
\textrm{d}N/\textrm{d}\Omega \approx \frac{1}{\sqrt{2\pi}\Theta_0}\exp (-\Theta^2 / 2 \Theta_0^2 )~,
\end{equation}
where \(\Theta_0\) is given by
\begin{equation}
\Theta_0 = \frac{13.6\UMeV}{\beta pc}\sqrt{x/X_0}[1+0.038\ln (x/X_0)]~,
\end{equation}
with \(X_0 \) defined as
\begin{equation}
X_0 = \frac{716.4\Ug\Ucm^{-2}\UA}{Z/(Z+1) \ln (287 / \sqrt{Z} )}~.
\end{equation}

Multiple scattering of the protons as they traverse the sample degrades the spatial resolution pos\-sible from ideal energy loss imaging, but it can also increase the contrast and hence the thickness reso\-lution.  This is because those protons scattered away from the direct line of sight from the source to the film detector are moved from the umbral to penumbral region on the film, thus reducing the flux of protons in the direct shadow.  The decrease in proton flux associated with a given film plane being sensitive to higher initial energy protons, owing to the energy loss in the sample, is augmented by a flux reduction from scattering.  The very small-angle scattered protons, however, increase the net flux of protons in the penumbral region, which can lead to `limb brightening' effects, which are usual for image techniques based on scattering (rather than absorption). \\

\subsubsection{Field deflection}

Probably the most important applications to date of proton probing are related to the unique capability of this technique to detect electrostatic fields in plasmas \cite{borghesi2002a, borghesi2003}. This has allowed researchers to obtain, for the first time, direct information on electric fields arising through a number of laser--plasma interaction processes \cite{mackinnon2004,borghesi2003a, borghesi2002}. The high temporal resolution is here fundamental in allowing the detection of highly transient fields following short-pulse interaction.
When the protons cross a region with a non-zero electric field, they are deflected by the transverse component \(E_\perp \) of the field. The proton transverse deflection at the proton detector plane is equal to
\begin{equation}
\Delta r_\perp \approx eL \int_0^b (E_\perp /m_\textrm{p} v_\textrm{p}^2 )\textrm{d}l~,
\end{equation}
where \(m_\textrm{p} v_\textrm{p}^2 /2 \) is the proton kinetic energy, \(e\) its charge, \(b\) the distance over which the field is present, and \(L\) the distance from the object to the detector.
As a consequence of the deflections, the proton beam cross-section profile undergoes variations showing local modulations in the proton density. Assuming the proton density modulation to be small \(\delta n/n_0 \ll 1\), where \(n_0\) and \(\delta n\) are, respectively, the unperturbed proton density and proton density modulation at the detector plane, we obtain \( \delta n / n_0 \approx - \textrm{div} (\Delta r_\perp )/M\), where $M$ is the geometrical magnification. The value of the electric field amplitude and the spatial scale can then be determined if a given functional dependence of $E$ can be inferred a priori, \eg from theoretical or geometrical considerations. More details can be found in the references cited in this paper and in Ref. \cite{borghesi2006}, from where a part of this section was taken.

\subsection{Warm dense matter}
\label{subsec:Warm Dense Matter}

The creation of extreme states of matter is important for the understanding of the physics covered in vari\-ous research fields, such as high-pressure physics, applied material studies, planetary science, inertial fusion energy, and all forms of plasma generation from solids. The primary difficulties in the study of these states of matter are that the time-scales or the changes are rapid (\({\approx}1\Ups\)), while the matter is very dense and the temperatures are relatively low, \({\sim}\UeVZ/k_\textrm{B}\). With these parameters, the plasma exhibits long- and short-range orders, which are due to the correlating effects of the ions and electrons. The state of matter is too dense or too cold (or both) to admit standard solutions used in plasma physics. Perturbative approaches using expansions in small parameters for the description of the plasma are no longer valid, proving a tremendous challenge for theoretical models. This region where condensed matter physics and plasma physics converge is the so-called warm dense matter regime \cite{lee2003}.

Warm dense matter conditions can be generated in a number of ways, such as by laser-generated shocks \cite{saiz2008} or laser-generated X-rays \cite{glenzer2003, glenzer2007}, intense ion beams from conventional accelerators \cite{tahir2008}, or laser-accelerated protons \cite{patel2003, dyer2008, antici2006}, to name just a few. Whereas lasers only interact with the surface of a sample, ions can penetrate deeply into the material of interest, thereby generating large samples of homogeneously heated matter. The short pulse duration of intense ion beams furthermore allows for the investigation of the equation of states close to the solid-state density, because of the material's inertia, which prevents the expansion of the sample within the interaction. Moreover, the interaction of ions with matter is dominantly due to collisions and does not include a high-temperature plasma corona as is the case in laser--matter interaction.

The generation of large, homogeneous samples of warm dense matter is accompanied by the chal\-lenging task of diagnosing this state of matter, as the usual diagnostic techniques fail under these con\-ditions. The material density results in a huge opacity and the relatively low temperature does not allow traditional spectroscopic methods to be applied. Moreover, the sample size, deposited energy and life\-time of the matter state are strongly interrelated and are dominated by the stagnation time of the atoms in the probe. Thus, high spatial and temporal resolution is required to gain quantitative data in those experiments. Owing to the high density of the sample, laser diagnostics cannot be used. The properties of matter could be determined by measuring the expansion after the heating \cite{dyer2008} or by measuring the thermal radiation emitted by the sample \cite{patel2003}. However, even more interesting are the plasma parameters deep inside the sample, where the ion heating is most effective. An ideally suited diagnostic technique recently developed is X-ray Thomson scattering \cite{glenzer2003, landen2001, schollmeier2006}.

\begin{figure}
        \centering
        \includegraphics[width=0.8\textwidth]{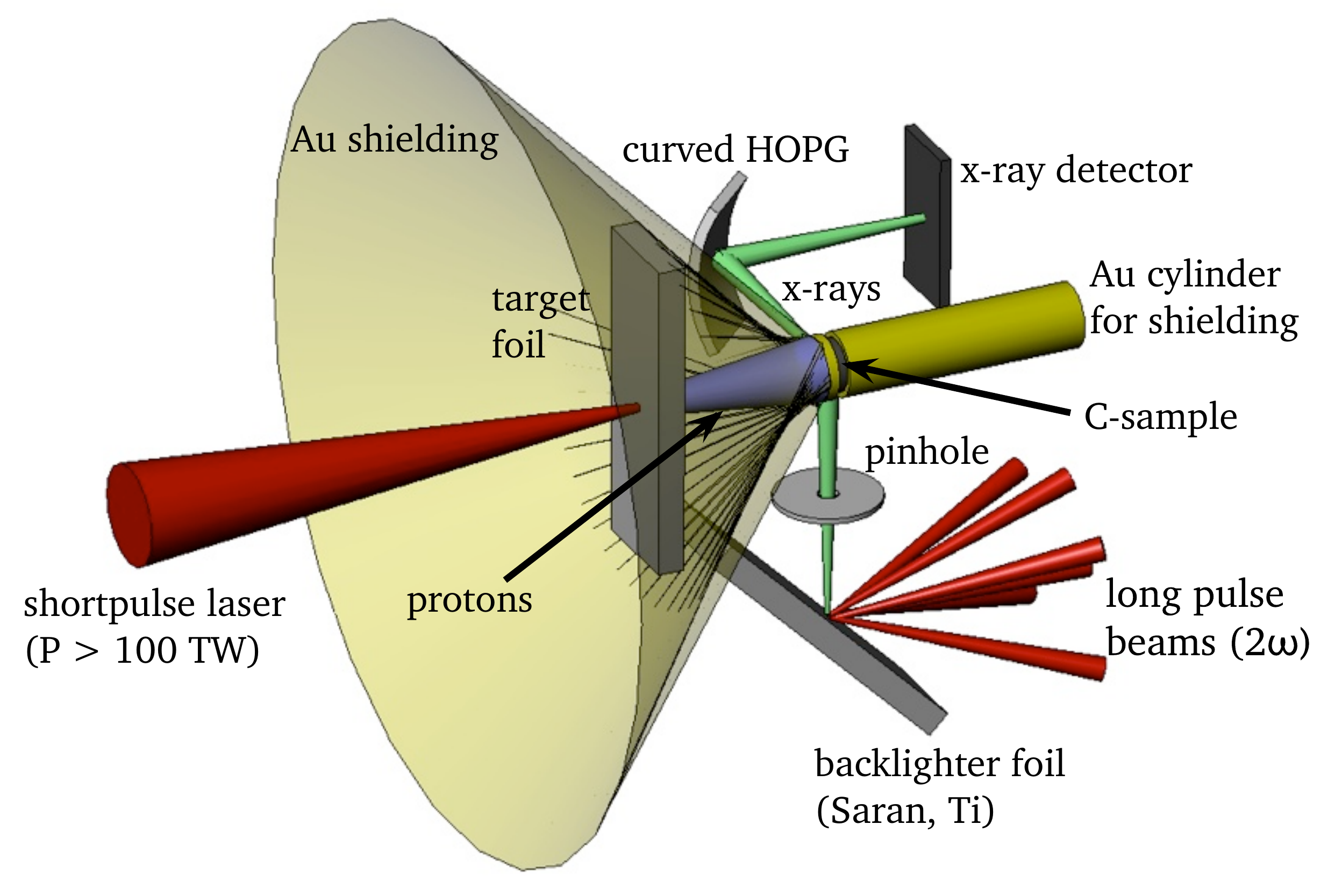}
                \caption{Experimental scheme to investigate the properties of laser-accelerated proton-heated matter by spectrally resolved X-ray Thomson scattering. HOPG, highly ordered pyrolytic graphite.}
                \label{wdmsetup}
\end{figure}

\Figure[b]~\ref{wdmsetup} shows a typical experimental scheme to investigate the transformation of solid, low-\(Z\) material into the warm dense matter state. The experimental scheme requires a high-energy short-pulse laser and one or more long-pulse laser beams in the same experimental vacuum chamber. In recent years, more and more laser facilities have upgraded their laser systems for such types of pump-probe experiment. A chirped pulse amplification laser beam above \(100\UW[T]\) power generates an intense proton beam from a thin target foil. The protons hit a solid-density sample and heat it isochorically, up to a temperature of several \(\UeV/k_\textrm{B}\). The long-pulse beams are used to drive an intense X-ray source from a titanium or Saran (which contains chlorine) foil. The sample is probed by narrow-band line-radiation from the chlorine or titanium plasma. The scattered radiation is first spectrally dispersed by a highly efficient, highly oriented, pyrolytic graphite crystal spectrometer with von Hamos geometry before it is detected. Extensive gold shielding (partly shown in \Fref{wdmsetup}) is required, to prevent parasitic signals in the scatter spectrometer. From the measured Doppler-broadened, Compton-downshifted signal, the temperature and density can be inferred.

Whereas lasers only interact with the surface of a sample, ions can penetrate deeply into the material of interest, thereby generating large samples of homogeneously heated matter. The short pulse duration of laser-produced ion beams furthermore allows for the investigation of equation of states close to the solid-state density, because the material's inertia prevents expansion of the sample during the interaction. In addition to these unique characteristics, the interaction of ions with matter is dominantly due to collisions and does not include a high-temperature plasma corona, as it is present in laser--matter interaction. The absence of a large radiation background is of importance to the experiment. Large conversion efficiencies have been observed, and significant energy can be transferred from the ultra-intense laser via the ion beam into the sample of interest. Because of the high beam quality, ballistic focusing has been demonstrated, allowing for an increase in local energy deposition, thus leading to higher temperatures. The use of hemispherical targets, including cone-guided targets to enhance the local proton flux on the material of interest, can even enhance the locally deposited energy.

\begin{figure}
        \centering
        \includegraphics[width=0.75\textwidth]{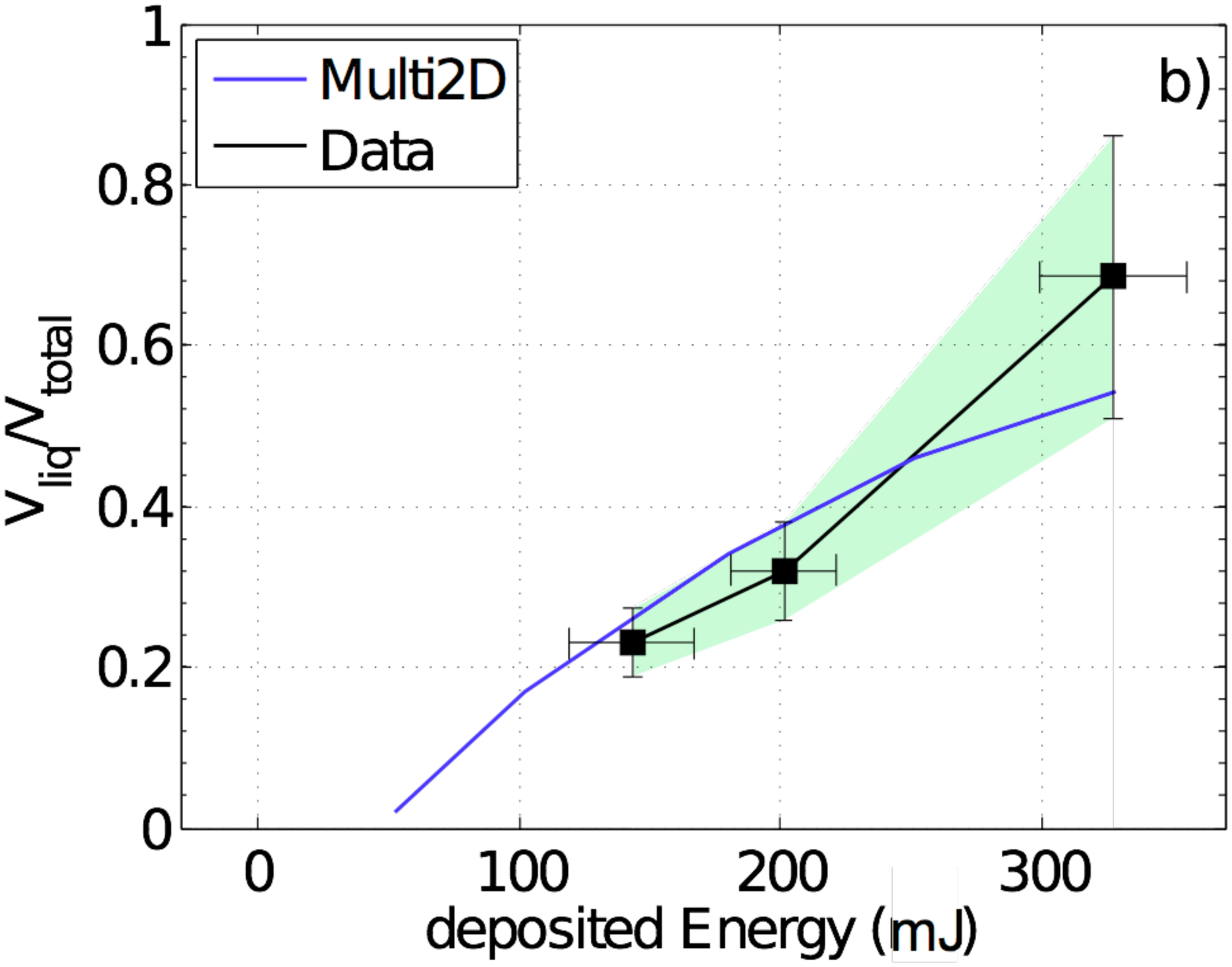}
                \caption{Molten fraction of a carbon sample versus deposited energy by the proton beam, calculated from the experimental data and simulated with Multi2D \cite{pelka2010}.}
                \label{moltencarbon}
\end{figure}

Using laser pulses in excess of \(100\UJ\), the intense proton beams can heat large targets up to several times the melting temperature. In a milestone experiment at TAW last year,  the molten fraction in carbon samples heated by intense proton beams was measured \cite{pelka2010}. \Figure[b]~\ref{moltencarbon} shows some results compared with a radiation hydrodynamics simulation, which uses SESAME as the equation-of-state model. It can be seen that agreement is better at lower temperatures, where ionization is not important, but at higher temperatures, the presence of an ionic component may be important.

To summarize, laser-accelerated proton beams are very well suited to produce macroscopic samples of warm dense matter. Their unique feature, having pulse durations of only a few picoseconds while containing more than \(10^{12}\) protons, cannot be matched by conventional ion accelerators.

\subsection{Fast ignition}
\label{subsec:Fast ignition}

In conventional inertial fusion, ignition and propagating burn occurs when a sufficient tempera\-ture (5--\(10\UkeV\)) is reached within a sufficient mass of deuterium-tritium fuel characterized by a density-radius product greater than the range of an alpha particle, \((\rho r)_\alpha > 0.3\Ug\Ucm^{-2}\). The necessary conditions for propagating deuterium-tritium burn are achieved by an appropriate balance between the energy gain mechanisms and the energy loss mechanisms. Mechanical work (\(P\textrm{d}V\)), alpha particle energy de\-position and, to a smaller extent, neutron energy deposition are the principal energy gain mechanisms in deuterium-tritium fuel. Electron heat conduction and radiation are the principal energy loss mech\-anisms. When the rate of energy gain exceeds the rate of energy loss for a sufficient period of time, ignition occurs. Because of the short burn time and the inertia of the fuel, the contribution of expansion losses is negligible.
Fast ignition \cite{tabak1994, temporal2002} was proposed as a means to increase the gain, reduce the driver energy and relax the symmetry requirements for compression, primarily in direct-drive inertial confinement fusion. The concept is to precompress the cold fuel and subsequently to ignite it with a separate short-pulse high-intensity laser or particle (electron or ion) pulse. Fast ignition is being exten\-sively studied by many groups worldwide, using short-pulse lasers or temporally compressed heavy-ion beams. There are several technical challenges for the success of laser-driven fast ignition. Absorption of the igniter pulse generates copious relativistic electrons, but it is not yet known whether these electrons will propagate as a stable beam into the compressed fuel to deposit their energy in a small volume.
In principle, heavy-ion beams can have advantages for fast ignition. A focused ion beam may maintain an almost straight tra\-jectory while traversing the coronal plasma and compressed target, and ions have an excellent coupling efficiency to the fuel and deliver their energy in a well-defined volume, owing to the higher energy de\-position at the end of their range (Bragg peak) \cite{caruso1996}. With the discovery of TNSA ions with excellent beam quality, the idea of using those beams for fast ignition was introduced \cite{roth2001, ruhl2001}. Protons do have several advantages compared with other ion species \cite{bychenkov2001} and electrons. First, because of their highest charge-to-mass ratio, they are accelerated most efficiently up to the highest energies. They can penetrate a target more deeply to reach the high-density region, where the hot spot is to be formed, because of the quadratic dependence of the stopping power on the charge state. And finally, they do, like all ions, exhibit a characteristic maximum of the energy deposition at the end of their range, which is desirable, to heat a localized volume efficiently.

The basic idea is to use a number of short-pulse lasers to irradiate a thin foil. The protons were accelerated from the rear surface of the foils and, because of the parabolic geometry, are focused into the compressed fuel. One of the requirements for proton fast ignition is the possibility of focusing the proton beams into a small volume. It has recently been demonstrated that proton beam focusing is indeed possible, and spot sizes of about \(40\Uum\) have been achieved. This is comparable to what is required by proton fast ignition. Larger irradiated areas on the target front surface, as required for proton fast ignition, would flatten the electron distribution at the rear surface. Not only might this result in a single divergence angle for different energies, but it would also form a much smaller initial divergence angle that could be compensated, in order to reach the required focal spot diameters.

\begin{figure}
        \centering
        \includegraphics[width=0.75\textwidth]{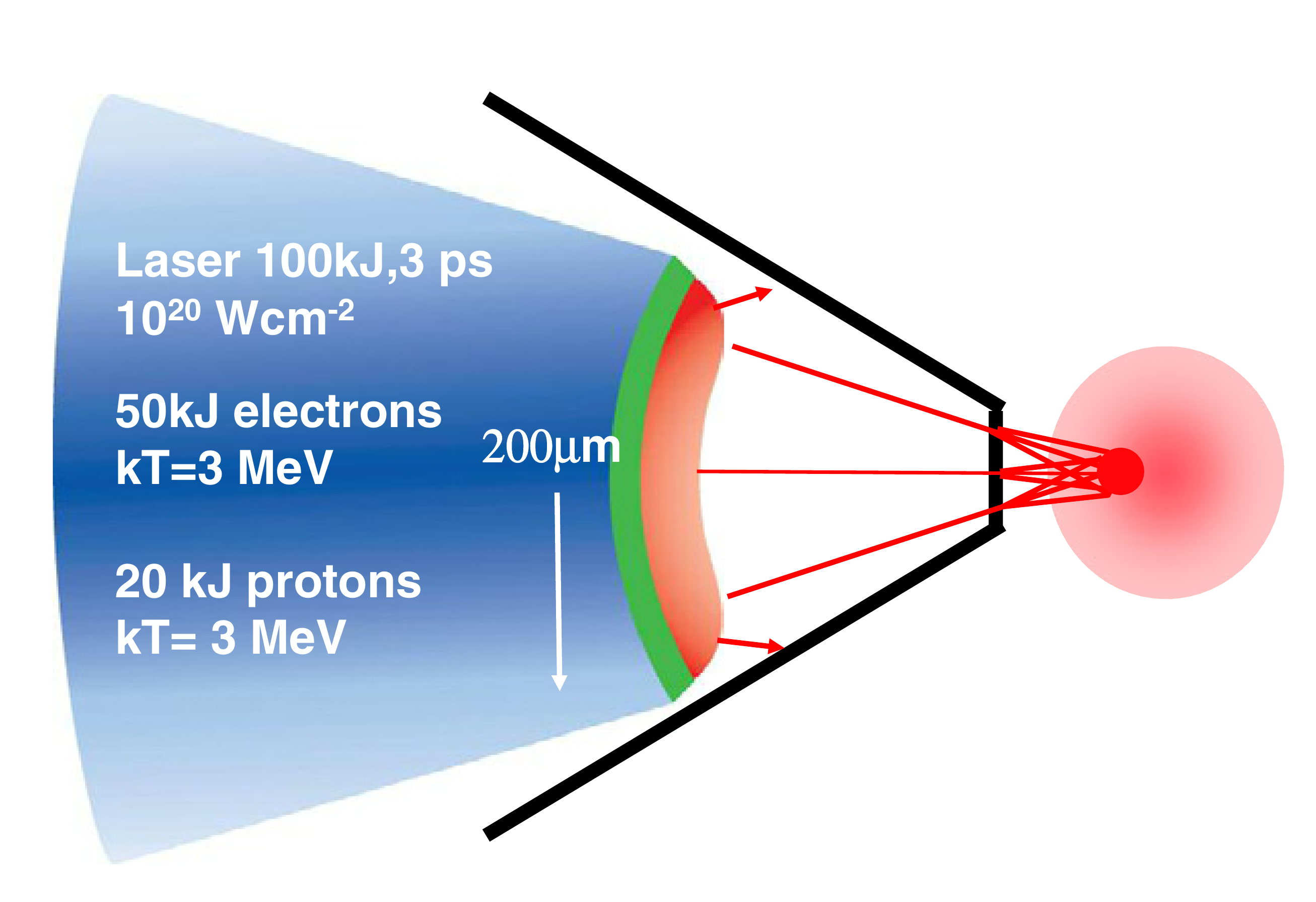}
                \caption{Proposed concept of using cone-guided proton fast ignition. Courtesy of M. Key (LLNL) }
                \label{pfi}
\end{figure}

The pulse length at the source is of the right order of magnitude for proton fast ignition, which has already been indicated in first experiments on ion acceleration. The protons are not monochromatic but rather have an exponential energy distribution. This seemed to be a concern at the beginning because of dispersion and pulse lengthening. A close distance to the pellet, however, raises concerns as to whether the thin metallic foil, which is to be the source of the protons, can be kept cold enough that it does not develop a density gradient at the rear surface, which would diminish the accelerating field. A second concern was related to the stopping power. Because of the difference in initial velocity, the energy deposition of protons with different kinetic energies is spread over a larger volume. Slower protons are stopped at a shorter distance and do not contribute to the creation of the igniter spark.
Fortunately, further work has relieved those concerns. Simulations by Basko \textit{et al.} (presented at the Fast Ignition Workshop 2002, Tampa, Florida) have shown that the protective shield placed in front of the source can withstand the X-ray flux of the pellet compression and keep the rear surface of the source foil cold enough for acceleration via TNSA. A thickness of a few tens of micrometres, however, provides thermal shielding as well as sufficient mechanical stability. The distance between the source and the ignition spot can be a few millimetres. If this distance is too short for the compression geometry (\eg if a closely coupled hohlraum is not used), the distance can be adjusted using a similar cone target, as for conventional fast ignition. As for the concerns on the hydrodynamic stability of the proton source foil, the proposed usage of a cone target similar to the one proposed for electron fast ignition \cite{norreys2000, kodama2001} has solved most of the problems, by shielding the foil from primary soft X-rays generated in the compression of the capsule (see \Fref{pfi}). Furthermore it was demonstrated that small-scale plasma density gradients at the rear side of the proton source target caused by target preheating have no significant (less than 10\%) impact on the TNSA mechanism \cite{fuchs2007}.

A big surprise was the fact that a monochromatic proton beam is actually not the optimum to heat a hot spot in a fusion target. Numerical simulations have shown that one has to take into account the decrease of the stopping power of the nuclear fuel with increasing plasma temperature. Thus, an exponential energy spectrum, like the one generated by this mechanism, is the most favourable. The first protons with the highest energies penetrate the fuel deeply. By the time the proton number increases and the target temperature rises, the stopping power is reduced, thereby compensating for the lower initial energy of the incoming protons. Thus the majority of the protons deposit their energy within the same volume.

Existing short-pulse lasers have already demonstrated intensities that are sufficient for gener\-ating the proton energy spectra required for proton fast ignition. Regardless of the nature of the igniter beam, calculations show a minimum deposited energy required for fast ignition of the order of a few tens of kilojoules. There have been many experiments with different laser systems accelerating proton beams. Interestingly, the laser to ion beam conversion efficiency seems to increase strongly with total laser energy from a fraction of a per cent up to more than 10\%. Carefully extrapolating the conversion efficiencies to multi\-kilo\-joule laser systems, conversion efficiencies of more than 10\% can be expected, which would result in the need for a few hundred kilojoules of short-pulse laser energy for proton fast ignition.

The most detailed theoretical analysis so far has been published by Temporal \textit{et al.} \cite{temporal2002} for a proton beam with an exponential energy spectrum. Following their assumptions, a total proton energy of about \(26\UkJ\) at an effective temperature of \(3\UMeV\) is required. This moderate temperature was found to be an optimum between the need for high temperatures to minimize the pulse lengthening caused by the velocity spread and the stopping range for the major part of the spectrum. It is interesting to note that the protons, which effectively heat the hot spot, contain only \(10\UkJ\) of the total energy and range from 19 to \(10\UMeV\). If it could be possible to shape the energy spectrum of the laser-accelerated protons, this would strongly influence the required laser beam energy. The total number of protons needed for ignition is close to \(10^{16}\). Is it conceivable to achieve a consistent scenario for those requirements? A typical proton beam temperature of \(3\UMeV\) is commonly obtained in experiments at \(5 \times 10^{19}\UW\Ucm^{-2}\). Assuming a pulse length of \(4\Ups\) (which would increase the damage threshold of modern dielectric compressor gratings) and a conversion efficiency of 10\%, a total laser energy of \(260\UkJ\) would be needed.

The use of a cone-guided geometry, as in conventional fast ignition, has been considered to be of great advantage. The source foil can be shielded from the radiation caused by the primary drivers, the source-to-hot spot distance can be tailored precisely, and the pellet can be protected from heat during the injection into the target chamber. A recent experimental campaign to study the influence of the cone walls on the propagation and the transport of TNSA protons has shown that, despite the influence of
self-generated electric fields in the cone walls by the recirculating electrons, good focusing may be achievable.

After the initial introduction of laser-accelerated proton beams for fast ignition, theoretical studies have not only quantified the required beam parameters \cite{temporal2002, caruso1996}, but also recently introduced sophisti\-cated scenarios that have greatly relaxed those parameters. A recent study proposed a combination of two spatially shaped proton beam pulses with a total beam energy that would match laser systems that might be available in the not too distant future \cite{temporal2006, temporal2008}. The most recent scenario looks for a ring-shaped proton beam, to impact into the dense fuel and further compress the hot spot, with a subsequent pulse of protons in the centre, to ignite the double-compressed core. This would cause an energy that is further reduced by a factor of two compared with the model described previously. Such laser proton beams have been generated using advanced cone geometries \cite{deppert2011}.

\subsection{Medical applications}
\label{subsec:Medical Applications}

Soon after the discovery of TNSA ion beams, the prospects for medical applications have been the focus of research. Besides the possibility of transmuting short-lived isotopes for positron emission tom\-ography, the main interest was in the use as a driver for hadron therapy. Hadron therapy  \cite{chu1993, khoroshkov1995, kraft2000, wilson1946, amaldi1999} is a radiotherapy technique that uses protons, neutrons, or carbon ions to irradiate cancer tumours. The use of protons and  carbon ions in radiotherapy  has several advantages over the more widely used X-radiotherapy. First, the proton beam scattering on the atomic electrons is weak and thus there is less irradiation of healthy tissues in the vicinity of the tumour. Second, the slowing down length for a proton of given energy is fixed, which avoids undesirable irradiation of healthy tissues at the rear side of the tumour. Third, the well localized maximum of proton energy losses in matter (the Bragg peak) leads to a substantial increase of the irradiation dose in the vicinity of the proton stopping point.

The proton energy window of therapeutical interest  ranges between 60 and \(250\UMeV\), depending on the location of the tumour. Proton beams with the required parameters are currently obtained using conventional charged-particle accelerators, such as synchrotrons, cyclotrons, and linacs \cite{wieszczycka2001}. The use of  laser-based accelerators  has been proposed as an alternative  \cite{bulanov2002, bulanov2002a, fourkal2003, fourkal2004, malka2004}, which could lead to advantages in terms of device compactness  and cost.

A laser accelerator could be used simply as a high-efficiency ion injector for the proton accelerator, or could replace a conventional proton accelerator altogether. Because of the broad energy and angular spectra of the protons, an energy selection and beam collimation system will be needed \cite{fourkal2002, fourkal2003a}. Typically, \(\Delta E / E \approx 10^{-2}\) is required for optimal dose delivery over the tumour region.  All-optical systems have also been proposed, in which the ion beam acceleration takes place in the treatment room itself and ion beam transport and delivery issues are minimized. In this case, the beam energy spread and divergence would need to be minimized by controlling the beam and target parameters.
The required energies for deep-seated tumours \(({>}200\UMeV)\) are still in the future, but appear to be within reach, considering ongoing developments in the field. A demanding requirement to be satisfied is also the system duty factor, \ie the fraction of  time during which the proton beam is available for use, which must not be less than \(0.3\).

With the recent experimental results of ion beams in the range up to \(80\UMeV\), the lower threshold for medical applications has been achieved. However, for deep-seated tumours it is uncertain whether the TNSA mechanism is still the best option or whether new mechanisms should be explored that not only lead to higher particle energies, but also offer a much smaller energy dispersion to begin with.

\subsection{Neutron source}
\label{subsec:Neutron Source}

Since the first experiments with ultra-intense lasers, nuclear reactions have been observed and also used to diagnose the hot core part of the laser--plasma interaction \cite{guenther2011}.
In addition to the generation and detection of radio-isotopes and transmuted nuclei, neutrons are released either as a result of intense bremsstrahlung or by electron or ion impact. Because of the large conversion efficiency of laser to ion beam energy and the large cross-section for subsequent proton neutron reactions, laser-driven neutron sources based on the TNSA mechanism have become a focus of modern nuclear research.

One has to distinguish between the different neutron generation mechanisms. At proton beam particle energies in the megaelectronvolt range, the interaction and neutron generation relies on the exci\-tation of giant resonances that result in single (p, n) or multiple (p, \(\times\)n) neutron emission. The cross-section can be quite large and is energy dependent, peaking at characteristic proton impact energies. In the case of two particles in the exit channel, the neutron spectrum is monoenergetic for a given projectile energy and neutron emission angle. However, since the angle and energy spread of laser-emitted particles is large, only strongly exothermal reactions (\(Q \gg E_{\textrm{proj}}\)) will yield roughly monoenergetic neutrons. Which process takes place in a particular instance depends on the combination of target, projectile, and momentum transfer. The cross-sections for these processes are in the range of \(100\Umb\) up to \(1\Ub\) and are therefore quite large.

 As the driving ion beam is ultra-short and the release mechanism is prompt, the neutron pulses are very short and originate from a very small region maximizing the net flux on secondary samples. Such a probe exists in the form of fusion neutrons. They are generated by the d(d,n)\(^3\)He fusion reaction in deuterated targets, and their use as a laser-plasma diagnostic is not fundamentally new. When neutrons are produced from laser-accelerated ions in the bulk of an irradiated \((\EC\ED_2)_n\) target, they are emitted within a few picoseconds from a volume of the order of \((10\Uum)^3\). During the neutron pulse, at a distance of several millimetres from the target, fast neutron fluxes of \(10^{19}/(\UcmZ^2\cdot\UsZ)\) can be achieved, which is four orders of magnitude higher than current continuous research reactors can deliver.

In the past, the neutron emission caused by (\(\gamma\), n) and (p, n) reactions from the target have been measured at moderate laser intensities. A typical detector set-up is a silver activation detector attached to a photomultiplier tube. On typical shots, the neutrons are generated by (\(\gamma\), n) reactions within the target (caused by the bremsstrahlung photons from the relativistic electrons) and by (p, n) reactions of our proton beam impacting on the radiochromic
film screen or a dedicated secondary production target. This can be, \eg a target of deuterated plastic (\(\EC\ED_2\)), which was irradiated by a beam of TNSA-accelerated deuterons. One can observe the yield of neutrons from (d, d) fusion reactions.

To optimize laser-driven neutron sources, one can perform simulation studies using con\-soli\-dated findings on particle beam characteristics obtained from laser experiments \cite{guenther2011a}. The optimization will be according to the absolute neutron yield and angle, as well as the spectral distribution. For neutron gener\-ation, we attempt a two-stage target design in which the TNSA ion beam irradiates a secondary sample. The advantage of this design is that we can optimize the proton or deuteron generation using different targets in the first stage. According to the beam properties obtained from the first stage, it will be possible to optimize the neutron production via proton- and deuteron-induced neutron disintegration reactions, respectively, in the second stage. The neutron production target design (second stage) allows adaptation to the desired application.
\begin{figure}
        \centering
        \includegraphics[width=0.75\textwidth]{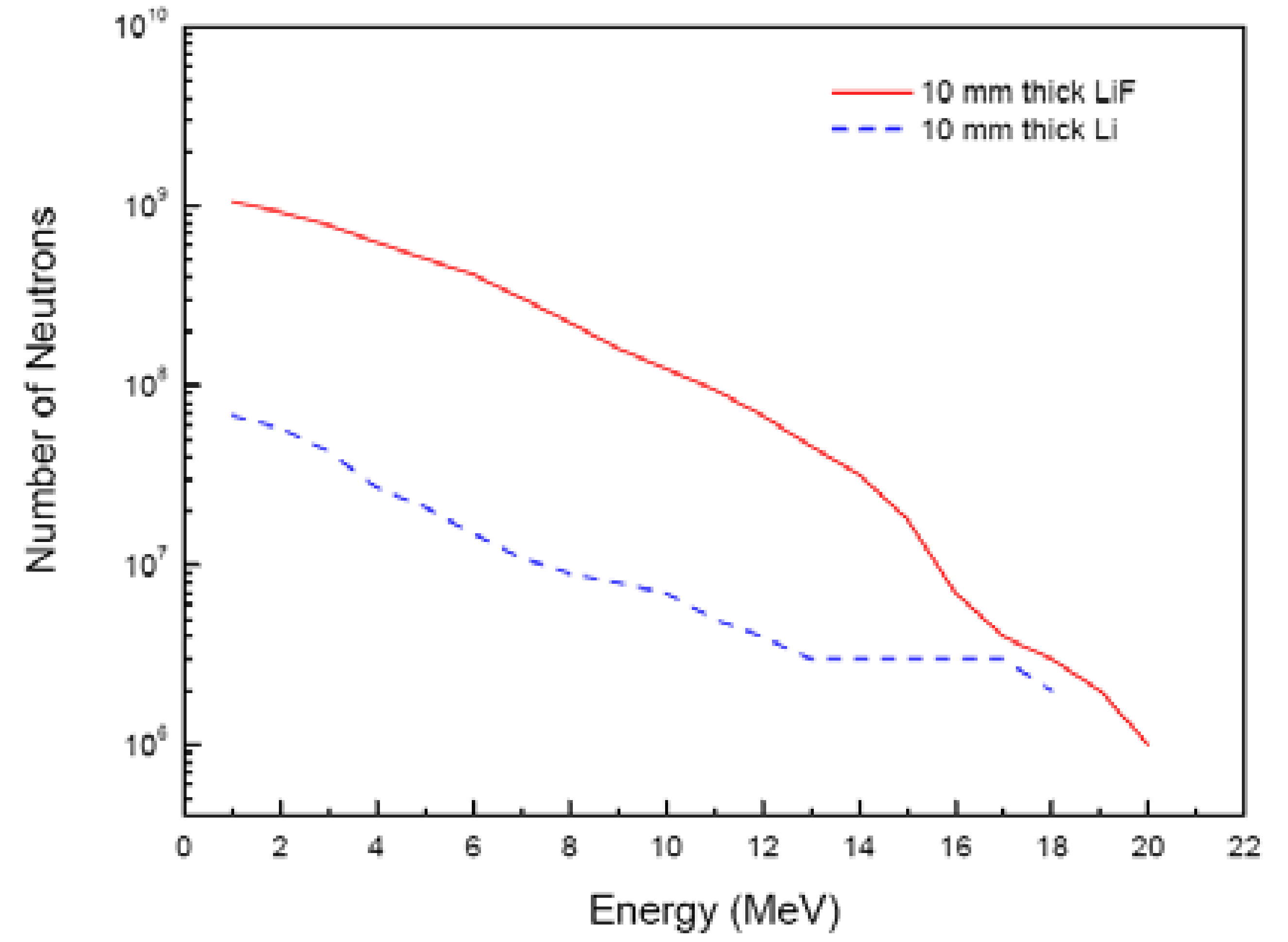}
                \caption{Simulated neutron spectra for different converter materials}
                \label{neutspec}
\end{figure}
In earlier experimental campaigns at the PHELIX laser facility at the GSI Helmholtz Centre for Heavy-Ion Research, \({>}10^9\) neutrons per shot from proton-induced reactions in copper have been produced. The integrated number of protons was \(10^{12}\) to \(10^{13}\). Each neutron yield in these experiments already exceeded that from the accelerator-driven neutron source FRANZ \cite{ratzinger2010} in Frankfurt (Germany) by five orders of magnitude.
With the help of the GEANT4 code \cite{agostinelli2003}, we can simulate the expected neutron yield for TNSA protons using experimental input spectra.
We have simulated the neutron spectra and the production efficiencies using several isotopes within the second-stage target. The thickness of these different targets was \(10\Umm\). The highest production efficiencies were obtained from proton-induced reactions at lithium, beryllium, boron, and vanadium in their natural abundance.
As a benchmark, one can compare the simulation results for copper with experimental data,  where one finds a good agreement.

The neutron spectra from proton-induced reactions in beryllium and lithium show high particle numbers in the lower energy range and in the range around several megaelectronvolts, respectively. This is of interest in transmutation studies and nuclear material science. \Figure[b]~\ref{neutspec} demonstrates the difference in simulated neutron spectra using lithium in the natural abundance and lithium fluoride. The properties of the initial proton spectrum that was used in the simulation were obtained from experimental results at the PHELIX facility. The initial particle number was \(10^{13}\) and the maximum proton energy was \(25\UMeV\).

In addition, the neutron yield using lithium fluoride is much higher than the neutron yield from proton-induced reactions in lithium. The explanation is that the mass density of lithium fluoride is higher, owing to the interatomic compounds. Lithium fluoride has a mass density of \(2.64\Ug/\UcmZ^3\) and the mass density of lithium fluoride is only \(0.53\Ug/\UcmZ^3\). This demonstrates the attraction in using composite targets in future studies of laser-driven neutron sources. In future developments of the optimization of laser-driven neutron sources, we will use more sophisticated composite target designs for adaptation to the desired applications.

\end{document}